\documentclass[aip,jcp,amsmath,amssymb,floatfix,reprint,citeautoscript,longbibliography,noeprint]{revtex4-2}

\usepackage[utf8]{inputenc}
\usepackage[T1]{fontenc}
\usepackage{lmodern}

\usepackage{graphicx}
\usepackage{wrapfig}\usepackage[colorlinks,allcolors=black,citecolor=blue,urlcolor=blue]{hyperref}
\usepackage[mathlines,displaymath]{lineno}
\usepackage{placeins}
\usepackage{booktabs}
\usepackage{tikz}
\usetikzlibrary{decorations.pathreplacing,calligraphy}
\usetikzlibrary{shapes.misc, positioning}
\usepackage{tabularx}
\usepackage{xcolor}
\newcommand{\rev}[1]{{\color{black} #1}}

\DeclareUnicodeCharacter{2212}{-}
\graphicspath{{figures/}}

\begin{document}
\def\mytitle{%
Introduction to machine learning potentials for atomistic simulations
}
\title{\mytitle}
\author{Fabian L. Thiemann}%
\affiliation{%
IBM Research Europe, Daresbury, Warrington, WA4 4AD, UK
}
\affiliation{%
Cavendish Laboratory, Department of Physics, University of Cambridge, Cambridge, CB3 0HE, UK
}
\author{Niamh O'Neill}%
\affiliation{%
Yusuf Hamied Department of Chemistry, University of Cambridge, Lensfield Road, Cambridge, CB2 1EW, UK
}
\affiliation{%
Cavendish Laboratory, Department of Physics, University of Cambridge, Cambridge, CB3 0HE, UK
}
\affiliation{%
Lennard-Jones Centre, University of Cambridge, Trinity Ln, Cambridge, CB2 1TN, UK
}
\author{Venkat Kapil}%
\affiliation{%
Yusuf Hamied Department of Chemistry, University of Cambridge, Lensfield Road, Cambridge, CB2 1EW, UK
}
\affiliation{%
Lennard-Jones Centre, University of Cambridge, Trinity Ln, Cambridge, CB2 1TN, UK
}
\affiliation{%
Department of Physics and Astronomy, University College London, London, UK
}
\affiliation{%
Thomas Young Centre and London Centre for Nanotechnology, London, UK,  London, UK
}
\author{Angelos Michaelides}%
\affiliation{%
Yusuf Hamied Department of Chemistry, University of Cambridge, Lensfield Road, Cambridge, CB2 1EW, UK
}
\affiliation{%
Lennard-Jones Centre, University of Cambridge, Trinity Ln, Cambridge, CB2 1TN, UK
}
\author{Christoph Schran}%
\email{cs2121@cam.ac.uk}
\affiliation{%
Cavendish Laboratory, Department of Physics, University of Cambridge, Cambridge, CB3 0HE, UK
}
\affiliation{%
Lennard-Jones Centre, University of Cambridge, Trinity Ln, Cambridge, CB2 1TN, UK
}

\date{\today}

\begin{abstract}
Machine learning potentials have revolutionised the field of atomistic simulations in recent years and are becoming a mainstay in the toolbox of computational scientists.
This paper aims to provide an overview and introduction into machine learning potentials and their practical application to scientific problems.
We provide a systematic guide for developing machine learning potentials, reviewing chemical descriptors, regression models, data generation and validation approaches.
We begin with an emphasis on the earlier generation of models, such as high-dimensional neural network potentials (HD-NNPs) and Gaussian approximation potential (GAP), to provide historical perspective and guide the reader towards the understanding of recent developments, which are discussed in detail thereafter.
Furthermore, we refer to relevant expert reviews, open-source software, and practical examples -- further lowering the barrier to exploring these methods.
The paper ends with selected showcase examples, highlighting the capabilities of machine learning potentials and how they can be applied to push the boundaries in atomistic simulations.
\end{abstract}
{\maketitle}
\renewcommand{\tocname}{\vspace*{-2em}}
\tableofcontents

\section{Introduction}
\label{sec:intro}
Most of the chemistry and physics of molecular systems and materials
is governed by the potential energy surface (PES).
Within the Born-Oppenheimer approximation, the properties
of a system of interest can thus be obtained from
its thermally weighted population on the ground state PES,
as sampled either by molecular dynamics or Monte Carlo techniques.
Having access to an accurate but efficient representation
of the system's PES is therefore of paramount importance
for the computational study of material properties, reactions,
and molecular processes.
While \textit{ab initio} techniques such as density functional theory (DFT)
can provide the required accuracy for a large variety of complex systems, they are usually
relatively expensive as the electronic structure of each sampled
configuration needs to be obtained.
This cost comes from the challenges associated with approximating the
many-body Schrödinger equation, in particular, due to the electron-electron
repulsion.
Force field techniques,
on the other hand, use a set of usually physically motivated functions for different
types of interaction to represent the PES with
parameters optimized to either match experiments or higher-level
electronic structure data.
These are usually quite efficient but in many cases not accurate enough or
are missing reactivity in order to provide reliable insight.
In recent times, the use of machine learning has enabled the
PES to be learned from the previously solved electronic structure of a set of configurations
in order to provide reliable interpolation at a cost similar to
FF methods, but reproducing the accuracy of \textit{ab initio} techniques.~\cite{%
Behler2016/10.1063/1.4966192,%
Behler2017/10.1002/ANIE.201703114,%
Bartok2017/10.1126/SCIADV.1701816,%
Butler2018/10.1038/s41586-018-0337-2,%
Deringer2019/10.1002/ADMA.201902765,%
Kang2020/10.1021/ACS.ACCOUNTS.0C00472,%
Noe2020/10.1146/ANNUREV-PHYSCHEM-042018-052331,
Mueller2020/10.1063/1.5126336,
Ceriotti2021,%
Behler2021/10.1140/EPJB/S10051-021-00156-1,
Unke2021/10.1021/ACS.CHEMREV.0C01111,
Kocer2022/10.1146/ANNUREV-PHYSCHEM-082720-034254}
The generality and data-driven nature of these approaches has led
to a surge in the use and development of machine-learning techniques for
atomistic simulations.
The field of machine learning potentials (MLPs) has grown quickly in the last couple of years after the
seminal works of Behler and Parrinello in 2007~\cite{Behler2007/10.1103/PhysRevLett.98.146401} using artificial neural networks and 
Bart\'ok and coworkers using Kernel-based approaches in 2010.~\cite{Bartok2010}
Nowadays, there is a wide variety of methods and just some examples include moment-tensor potentials \cite{Shapeev2016/10.1137/15M1054183, Novikov2020/10.1088/2632-2153/ABC9FE}, atomic cluster expansion \cite{Drautz2019/10.1103/PHYSREVB.99.014104}, spectral neighbour analysis potentials~\cite{Thompson2015/10.1016/J.JCP.2014.12.018, Wood2018/10.1063/1.5017641}, message-passing based neural networks \cite{Batatia2022/10.48550/arxiv.2206.07697, Unke2019/10.1021/ACS.JCTC.9B00181, Zubatyuk2019/10.1126/sciadv.aav6490,Batzner2021/10.1038/s41467-022-29939-5,pmlr-v139-schutt21a}
and deep learning methods.~\cite{Zhang2018/10.1103/PHYSREVLETT.120.143001, Wang2018/10.1016/J.CPC.2018.03.016, Schutt2018/10.1063/1.5019779}
Typically, MLPs are made up of two main components: an encoding strategy that represents the molecular structure (commonly referred to as descriptors) and a regression technique mapping the atomic configuration space to the PES.
These models can be trained on data generated by electronic structure calculations and
then used to make predictions for systems that are too large or
complex to be treated with such methods.
Another advantage of these machine learning approaches is that they can be easily
parallelized, which allows for efficient calculations on high-performance computing
platforms.~\cite{Singraber2019/10.1021/ACS.JCTC.8B00770,Chen2021/10.1063/5.0063880,Lu2021/10.1016/j.cpc.2020.107624,musaelian_learning_2023}
Overall, the use of machine learning in atomistic simulations is a vibrant area
of research that is greatly enhancing the accuracy and efficiency
of molecular and materials modelling.
The aim of this tutorial is to bridge the gap between the theoretical formalism of machine-learning methods and their application by providing a practical guide for those interested in implementing machine-learning-based methods in their research.
It is particularly targeted at those with some experience in molecular simulations but who want to understand more about the practicalities of training and applying MLPs to their system of interest.
To enable a broader understanding but also facilitate comparisons, in each section, we will first describe the general principles applicable to all approaches and then focus on specific methods, which can then be contrasted. 
In particular, the two most common flavours of MLPs, artificial neural networks and Gaussian process regression and their associated descriptors, will be a particular focus -- given their well-established software and literature available, they are very accessible to the ML beginner.
We start with the two first established variants in both of these categories, the so-called high-dimensional neural network potentials (HD-NNPs)~\cite{Behler2007/10.1103/PhysRevLett.98.146401} and Gaussian approximation potentials (GAPs)~\cite{Bartok2010}, as a gateway for more recent approaches discussed afterwards.
The motivation for this is that core ideas remain mostly constant, and new
developments thus can be best understood in their historical context.
For the tutorial to be of direct practical use, we also compile in Tab.~\ref{tab:code} of the supporting information, some of the open-source software available for various tasks
and provide a Colab tutorial that exemplifies all the steps
of the development process of an MLP.

The field has experienced a significant surge in recent years, leading to an abundance of expert reviews on diverse aspects of MLPs authored by renowned pioneers.
This includes multiple excellent reviews on the HD-NNP~\cite{%
Behler2011/10.1039/C1CP21668F,
Behler2014/10.1088/0953-8984/26/18/183001,
Behler2015/10.1002/qua.24890,
Behler2021/10.1021/acs.chemrev.0c00868}
and GAP~\cite{Bartok2015/10.1002/QUA.24927,
Deringer2021} formalism,
as well as other broader reviews~\cite{%
Deringer2019/10.1002/ADMA.201902765,
Schleder2019/10.1088/2515-7639/AB084B,
Unke2021/10.1021/ACS.CHEMREV.0C01111,
Mishin2021/10.1016/J.ACTAMAT.2021.116980,
Zubatiuk2021/10.1021/ACS.ACCOUNTS.0C00868}.
In addition, special areas have been covered, such as structural representation by chemical descriptors~\cite{Musil2021/10.1021/ACS.CHEMREV.1C00021}, learning of excited states~\cite{Westermayr2021/10.1021/ACS.CHEMREV.0C00749}, machine learning in chemical compound space~\cite{Huang2021/10.1021/ACS.CHEMREV.0C01303}, focus on small molecules~\cite{Manzhos2021/10.1021/ACS.CHEMREV.0C00665} and reactions~\cite{Meuwly2021/10.1021/ACS.CHEMREV.1C00033}, inclusion of long-range effects~\cite{Anstine2022/10.1021/ACS.JPCA.2C06778}, as well as dataset generation techniques~\cite{Miksch2021}. 
Complementing this work, our aim is to present an entry-level description of the methodology, encompassing a broad part of this rapidly expanding field.
We hope this can serve as a solid foundation for further exploration using the aforementioned resources.
While the GAP and HD-NNP formalisms and later developments have all been
reviewed in detail before, they have not been directly contrasted in a
tutorial-style paper as an introduction for beginners.
Thus, this tutorial will function as a gateway to the more detailed and advanced material to facilitate further uptake of the methods in the everyday toolbox of computational scientists.

The basic idea behind machine learning in the current context is to assemble a set of
reference points for which the property is going
to be ``learned'' is known.
This is usually done in an automated procedure in order to
prevent as much user input as possible.
In the field of supervised learning, the properties (also called labels)
of the reference points need to be curated by the user before being learned by
the machine learning algorithm.
This preparation of the input is, in many cases, very important for the
quality of the outcome, since a clear differentiation between individual
points in the reference set needs to be achieved.
Once a meaningful data set has been prepared, a universal functional
form with large parameter sets is optimised to match the reference values (referred to as regression).
Afterwards, the model can be applied for fast and highly accurate
interpolation between the points in the reference set.

\begin{figure*}
    \centering{}
    \includegraphics[width=0.8\textwidth]{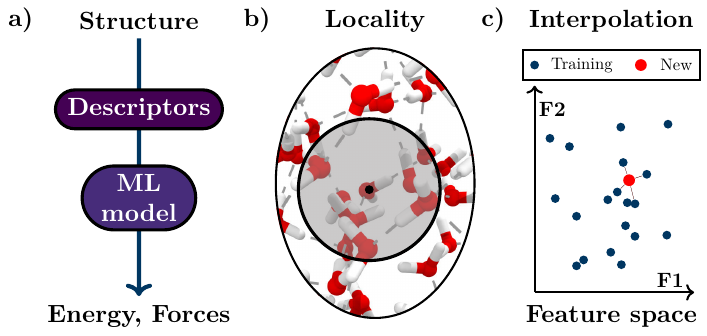}
    \caption{
        \textbf{Overview of the main principles of machine learning potentials.}
        a) Structure is related to energy and forces in a two-step process using chemical descriptors as input for the machine learning regression model.
        b) Most descriptors in the field are limited to a certain cutoff around each atom, building a locality approximation into the formalism as illustrated here by the grey-shaded region centred on the central oxygen atom.
        c) The main task of the machine learning potential is then, after training, to interpolate between the known and representative training data for previously unseen configurations. This is illustrated here within a simple two-dimensional feature space.
    }
    \label{fig:mlp-overview}
\end{figure*}

Within this beginner's guide, we will see how the concepts of machine learning
can be applied in atomistic simulations in order to ``learn''
efficient and reactive forms of the PES.
The main concepts shared by most approaches are summarised in 
Figure~\ref{fig:mlp-overview}.
In order to construct a representative structure-energy relation,
the structure is usually first transformed by a set of descriptors
to provide more meaningful input for the actual machine-learning model.
In most cases, these descriptors are atom-centred and limited to a certain spherical cutoff,
thus yielding a fingerprint of the chemical environment around each atom.
This means that locality is usually built into such models, enabling
linear scaling with the number of atoms.
The machine learning model is then trained to reproduce energies (and forces as
the first derivative of the PES) for a curated set of labeled configurations.
This is done by optimizing the parameters of the model in order to minimize
the difference between the model's prediction and the reference values.
Afterwards, the model is able to provide reliable predictions for unseen
configurations as long as sufficiently similar configurations have been
present in the training data.
We refer the novice reader to Tab.~\ref{tab:def} for a short overview of the central concepts of machine learning relevant for this tutorial.

In the next sections, we will go over the details of these various
steps in developing MLPs.
First, chemical descriptors will be introduced where the main focus will
be on radial and angular descriptors as well as atomic density.
Next, two of the most well-established and common machine learning approaches, artificial neural networks
and Gaussian process regression models will be presented, as well as how they can be combined with descriptors to provide robust and accurate structure-energy mappings.
We will then shift to strategies for the generation of representative data sets,
as well as validation techniques.
Finally, we end with an overview of showcase examples as well as an outlook on
new developments in the field and techniques, such as going beyond ground-state PES
and representing other properties.

\section{Chemical Descriptors}
\label{sec:descrip}
PESs have some intrinsic properties when it comes to their dependence
on structure:
Imagine translating, rotating, or reflecting a molecule.
All three operations do not influence the potential energy of the system,
as long as no external potential is applied.
This means that the potential energy is invariant with respect to
these three operations.
Similarly, the order of the atoms should not influence the energy either,
resulting in an invariance with respect to permutations.
Therefore, when constructing a mathematical relationship between structure and energy,
it is crucial to think about how the structural information is encoded
as input for the machine learning model.~\cite{Musil2021/10.1021/ACS.CHEMREV.1C00021, Langer2022/10.1038/s41524-022-00721-x, Zheng2021/10.1088/2632-2153/ABDAF7}
It is possible to construct a PES without considering these inherent invariances,
but it means that the data set needs to provide sufficient information on differently
rotated, translated, and permuted systems,
or the model itself must be able to recover the invariances from the input.
In such cases, Cartesian coordinates or similar measures like distance matrices,
collecting all $N \times N$ distances between all atoms in the system,
as well as their inverse (called Coulomb matrices)~\cite{Rupp2012/10.1103/PHYSREVLETT.108.058301},
are used to describe the chemical structure.
Early work captured the rotational and translational invariances using internal coordinate systems~\cite{Blank1998/10.1063/1.469597, Brown1998/10.1063/1.472596,Lorenz2004/10.1016/J.CPLETT.2004.07.076}. 
However, this still did not solve the issue of permutational invariance.
Usually, it is easier to incorporate the invariances by transforming
the Cartesian coordinates into a more representative form that intrinsically accounts for the required invariances.

In the following, we will see how two different types of descriptors
can be used to represent chemical environments and how the required
invariances are included.
First, so-called atom-centred symmetry functions are introduced,
which probe the radial and angular environment within a certain cutoff.
The second example of descriptors is the so-called smooth overlap
of atomic positions, relying on a density representation
of the atoms that are expanded as many-body expressions.
Finally, we will give a brief overview of the limitations
of these widely used descriptors and provide examples
of more modern variants.

\subsection{Atom-Centred Symmetry Functions}
\label{ACSF}

Atom-centred symmetry functions (ACSFs)~\cite{Behler2007/10.1103/PhysRevLett.98.146401,Behler2011/10.1063/1.3553717,Jose2012/10.1063/1.4712397} and variants~\cite{Gastegger2018/10.1063/1.5019667, Bircher2021/10.1088/2632-2153/ABF817}
have been introduced %
as a way to encode the chemical fingerprint
around each atom in a system while incorporating the translational,
rotational and permutational invariance.
They rely on a set of radial and angular functions that probe
different regions in the vicinity of each atom up to a predefined
cutoff.
For both the radial and angular symmetry functions, a cutoff function,
such as
\begin{align}
\label{eq:cut}
    f_\mathrm{c}(R_{ij})=
    \begin{cases}
        0.5\cdot \left[\cos\left(\frac{
        \pi R_{ij}}{R_\mathrm{c}}\right) + 1\right] \qquad &\text{for } R_{ij}\leq R_\mathrm{c}\\
        0  &\text{else}
    \end{cases}
\end{align}
is used. 
Here, $R_{ij}$ is the distance between the central atom $i$ and
a neighbouring atom $j$, used to define the atomic environment
up to a certain cutoff radius $R_\mathrm{c}$.
The radial arrangement of the atoms within this cutoff sphere is accounted for
by a product of a Gaussian and the cutoff function according to Eq.~(\ref{eq:cut}),
\begin{align}\label{radsymf}
    G^\text{rad}_i = \sum_j e^{-\eta(R_{ij}-R_\mathrm{s})^2}\cdot f_\mathrm{c}(R_{ij}) 
\enspace , 
\end{align}
where different regions around the central atom $i$ can be probed
by adapting the width of the Gaussian $\eta$ and the shifting parameter $R_\mathrm{s}$.
To complement the description of the environment around each atom, angular functions
of the form
\begin{align}
    G^\text{ang}_i = 2^{1-\zeta} \sum_{j,k\neq i,j \neq k} &(1+\lambda\cos\theta_{ijk})^{\zeta}\cdot e^{-\eta(R_{ij}^2+R_{ik}^2+R_{jk}^2)} \nonumber \\
    &\cdot f_\mathrm{c}(R_{ij})\cdot f_\mathrm{c}(R_{ik})\cdot f_\mathrm{c}(R_{jk})
\end{align}
are employed.
These depend on the angle $\theta_{ijk}$ between the central atom $i$ and
two neighbours $j$ and $k$, where $i,j$ and $k$ can be any atom 
of the system of interest.
Different angular regions are probed by adjusting the
exponent $\zeta$. 
The parameter $\lambda$, which can have values of $+1$ or $-1$,
is used to shift the maximum of the cosine either to $\pi$ or $2\pi$.

The different parameters ($R_\text{c}$, $\eta$, $R_\text{s}$, $\zeta$, $\lambda$) are
so-called hyperparameters of the model and need to be chosen by the user.
They can vary depending on the system of interest as different chemical
systems require different fingerprints for their chemical surrounding.
However, in practice, a well-chosen general set of functions can be applied
for various systems as long as all relevant regions around the atoms are
well represented.
The influence of the different hyperparameters on the shape of the different
symmetry functions is shown in Figure~\ref{fig:acsf-soap}.
The cutoff function smoothly decays to zero when reaching the chosen cutoff
value $R_\text{c}$.
The radial functions are Gaussian functions of variable width and centred at a chosen
position, thus allowing different radial regions around the central atom to be probed.
The functional form of the angular ACSFs is a bit more complex due to their angular
and radial contribution, but the shape of the angular contribution is sufficient
to understand most of its effects:
By choosing $\zeta$, the angular component can be made narrower or wider,
while $\lambda$ allows the centre of the angular width to be moved to a different
angular value.
With these parameters, different angular regions around each atom can be
distinguished.

Both radial and angular symmetry functions are evaluated over all distances
with atoms of a chosen element up to the cutoff and the resulting
value is summed up to give the fingerprint of this element.
This means that for each pair and triple of elements, a different
set of symmetry functions needs to be specified in order to distinguish
all possible chemical environments.
A set of functions is required in this case to provide sufficient sensitivity 
for different radial and angular regions.
For a simple example of a water molecule, one would require two sets
of symmetry functions, one for the oxygen atom and one for the hydrogen atoms.
The set of ACSFs for the oxygen atom then probes different radial distances
with respect to the surrounding hydrogen atoms.
This could be achieved, for example, by defining ten radial functions of various widths and shifted
to different distances $R_\text{s}$, where the exact number depends on the system of interest and width and shift.
If very narrow Gaussians are used, we would need a large set of radial functions to cover the entire cutoff region, while wider ones --- although less sensitive --- would cover the cutoff region with fewer functions.
Furthermore, the angular environment of the oxygen atom could be probed
with, e.g. four angular ACSFs that include both hydrogen distances from the central oxygen atom.
Again, the number of angular functions is up to the user and depends on the desired sensitivity to different angular regions around the atoms.
Other distances (OO) or angles (OHO, OOO) can be omitted as they are
not present for a single water molecule.
A minimal set of ACSFs for the hydrogen atoms would require
some radial functions for the HH and OH distances each,
as well as angular functions for the HHO triple.
Again, other angular functions for HOO and HHH can be ignored
because of the limited number of atoms of the water molecule.
\subsection{Smooth Overlap of Atomic Positions}
\label{sec:SOAP}

\begin{figure*}[t]
\centering
\includegraphics[width=0.85\textwidth]{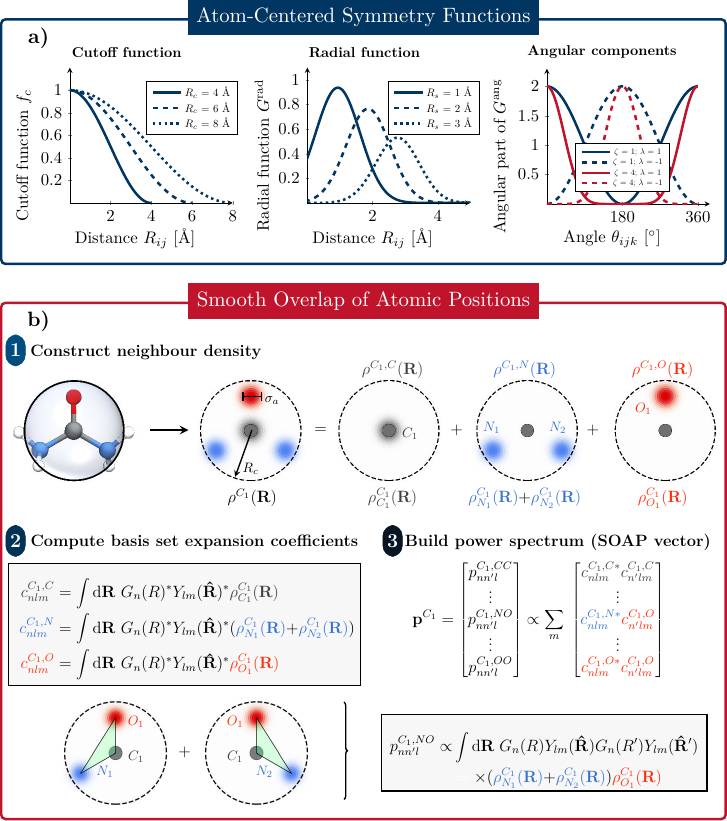}
\caption{\textbf{Summary of atom-centered symmetry functions (ACSF) and Smooth Overlap of Atomic Positions (SOAP).}
a) ACSF: Illustration of the three types of functions relevant for atom-centered symmetry functions.
        left) Cutoff function with three different cutoff radii.
        middle) Radial symmetry function with different shifted centers and $R_c$ = 6 \AA.
        right) Angular component of the angular symmetry function with different choices for $\zeta$ and $\lambda$.
b) SOAP: Schematic of the smooth overlap of atomic positions descriptor constructed for a non-planar urea molecule.
        First, atomic positions are transformed into the
        neighbour density $\rho$, which is permutationally invariant.
        Next, $\rho$ is expanded it in a local basis of radial
        functions and spherical harmonics, $Y_{lm}$.
        Finally, summing up the square modulus of the expansion
        coefficients $c_{nlm}$ over the index $m$ 
        provides the rotational invariance power spectrum $p$.
}
\label{fig:acsf-soap}
\end{figure*}

Developed by Bartók \textit{et al.}~\cite{Bartok2010,Bartok2013},
the smooth overlap of atomic positions (SOAP) provides an
alternative approach to atom-centred symmetry functions.
Aside from being a robust and invariant fingerprint of the atomic
environment, SOAP allows for the easy evaluation of the similarity
between two environments, making this descriptor particular 
powerful for kernel-based regression techniques introduced later.
To create the SOAP representation of the environment surrounding
a particular atom $i$, we follow the steps outlined in 
the bottom of figure \ref{fig:acsf-soap}.
First, we construct a set of neighbour densities,
$\rho^{i,s} (\textbf{R})$, one for each atomic species $s$
present in the system.
These densities are defined as sums of Gaussians with
variance $\sigma_{a}^2$ centred on all atoms of type $s$
within the neighbourhood of atom $i$:
\begin{equation}
    \rho^{i,s}(\textbf{R}) = \sum_{j} \delta_{ss_j} e^{-\frac{|\textbf{R}-\textbf{R}_{ij}|^2}{2 \sigma_\mathrm{a}^2}} f_\mathrm{c}(R_{ij})  \mbox{ , }
    \label{eq:SOAP_density}
\end{equation}
where the index $j$ runs over all neighbours of atom $i$, including itself,
within some cutoff distance $R_c$.
$\textbf{R}_{ij}$ is the vector pointing
from atom $i$ to the neighbour $j$, and $f_\mathrm{c}(R_{ij})$ is a cutoff function that ensures a smooth decay when approaching $R_c$
(as defined in Equation \ref{eq:cut}).
The Kronecker delta $\delta_{ss_j}$ ensures that only species of one desired
type are considered in each case.
The only hyperparameter at this stage aside from the cutoff distance, $R_c$,
is $\sigma_a$, which is often associated with the size of the
atoms \cite{Szlachta2014}
and determines the smoothness of the density.
Typically, $\sigma_a$ is set to around $0.3 \mathrm{\AA}$ 
when there are hydrogen atoms present in the system 
to account for the smaller X-H distances and
$0.5 \mathrm{\AA}$ when they are not \cite{Deringer2021}.

It is essential to understand that the complete set of elemental
neighbour densities are created for each atom $i$, regardless 
of the atomic number of atom $i$.
In a water molecule, for example, neighbour densities are always
constructed for both hydrogen and oxygen for each central atom,
thus resulting in a H and O neighbour density for both hydrogen atoms
as well as the oxygen atom.
By construction, these individual neighbour densities are invariant to
permutations within their respective element type.
To eventually ensure rotational invariance, they are expanded in a basis
of orthogonal radial functions $G_n(R)$ and spherical harmonics
$Y_{lm}(\mathbf{\hat{R}})$ such as
\begin{equation}
   \rho^{i,s}(\textbf{R}) \approx  \sum_{\substack{n<n_\mathrm{max} \\ l<l_\mathrm{max} \\ \left|m \right| \leq l }} c^{i,s}_{nlm} G_n(R) Y_{lm}(\mathbf{\hat{R}}) \mbox{ , }
  \label{eq:SOAP_density_harmonics}
\end{equation}
with $n$, $l$, and $m$ being integers while $l$ and $m$ are
known from quantum mechanics.
The choice of the radial basis is not really relevant to the outcome
of the procedure as long as it is sufficiently flexible.
Indeed, different bases have been used in the literature
from orthogonal polynomials to Gaussians.
In the given context, $n$ and $l$ are the indices for the
radial and angular channels, respectively,
and $\mathbf{\hat{R}}$ is the point on the unit sphere corresponding to
the direction of the vector $\mathbf{R}$.
In practice, the expansion is truncated at certain values for the
radial and angular expansion, represented by the hyperparameters
$n_\mathrm{max}$ and $l_\mathrm{max}$, respectively.
The related expansion coefficients $c^{i,s}_{nlm}$ are given by 
\begin{equation}
   c^{i,s}_{nlm}=  \int \mathrm{d}\mathbf{R}~G_n(R)^* Y_{lm}(\mathbf{\hat{R}})^* \rho^{i,s}(\textbf{R})
  \label{eq:SOAP_harmonics_coeff}
\end{equation}
from the inverse of equation~\ref{eq:SOAP_density_harmonics}.
Finally, the SOAP descriptor, also known as the SOAP vector,
is constructed using the so-called power spectrum of these coefficients
\begin{equation}
p^{i,ss'}_{nn'l}= \frac{1}{\sqrt{2l+1}} \sum_m (c^{i,s}_{nlm})^* c^{i,s'}_{n'lm} \mbox{ , }
\label{eq:SOAP_power_spectrum}
\end{equation}
where the sum over $m$ in the context of spherical harmonics corresponds
to averaging over all possible rotations, resulting in a rotationally
invariant representation.
This equation has a lot of indices and it is worthwhile to summarize
them again:
\begin{itemize}
    \item $i$ represents the central atom of choice for which we are obtaining the chemical fingerprint.
    \item $s$ and $s'$ represent different elements, thus providing information about different combinations of elements in the surroundings.
    \item Finally, $n$ and $n'$ represent different radial regions of these two element types.
\end{itemize}
Given this summation over different radial channels,
$p^{i,ss'}_{nn'l}$ encodes information about
pairs of vectors from the central atom and is therefore
of three-body nature.
The full SOAP descriptor characterising the environment around a given atom $i$,
$\textbf{p}^i \equiv \{ p^{i,ss'}_{nn'l}\}$, comprises all entries of 
$p^{i,ss'}_{nn'l}$ resulting in several hundreds or even thousands of components.
The length of this vector $N^\text{SOAP}$ scales quadratically with both 
the number of elements present in the system $N_\text{ele}$ 
and the radial expansion limit, $n_\mathrm{max}$, and
linearly with the angular expansion limit, $l_\mathrm{max}$
\begin{align}
    N^\text{SOAP} = \frac{n^{2}_\text{max}N^{2}_\text{ele}+n_\text{max}N_\text{ele}}{2} (l_\text{max}+1).
\end{align}
To get a grasp of the dimensions of the SOAP vector, let us
consider the descriptors used in previous work.
For instance, the SOAP descriptor is constructed to accurately describe
pristine graphene \cite{Rowe2018} used $n_\mathrm{max} = 8$ and
$l_\mathrm{max}=8$ resulting in a length of 324 elements. 
Employing identical expansion limits $n_\mathrm{max}$ and $l_\mathrm{max}$,
this value increases to 1,224 for hexagonal boron nitride being composed
of two elements B and N \cite{Thiemann2020}.
Given the increasing computational cost of calculating SOAP descriptors
for complex systems, it is important to carefully choose the hyperparameters
$n_\mathrm{max}$ and $l_\mathrm{max}$\cite{Rowe2020/10.1063/5.0005084,Rowe2022/10.1063/5.0091698/2841205} and several strategies
offer ways to construct a compressed version of the 
SOAP vector~\cite{Imbalzano2018/10.1063/1.5024611,Caro2019}.

While equation \ref{eq:SOAP_power_spectrum} provides an invariant
and robust representation of the atomic structure, the real power
of the SOAP descriptor lies in the convenience of constructing
a kernel function which can be used to measure the similarity
between two local environments.
This makes SOAP a popular choice for kernel-based regression
techniques such as Gaussian process regression.
The so-called SOAP kernel is calculated by taking the dot product
of the normalised power spectrum vectors, 
$\bm{\xi}^i = \textbf{p}^i / \sqrt{\textbf{p}^i \cdot \textbf{p}^i}$,
of the local environments.
For two environments, $A$ and $A'$, centred around atoms $i$ and $i'$,
respectively, the SOAP kernel is defined as
\begin{equation}
  k(A,A') = \left( \bm{\xi}^i \cdot \bm{\xi}^{i'} \right) ^ \zeta \mbox{ , }
  \label{eq:SOAP_kernel}
\end{equation}
where $k (A, A')$ takes values between 0 and 1 corresponding to full
dissimilarity or equality of the environments $A$ and $A'$, respectively.
The exponent $\zeta$ is an integer that defines the sharpness of
the kernel and needs to be $>1$ to produce a model beyond
three-body terms (see e.g. Ref~\citenum{Deringer2021} for a derivation).
In fact, many interatomic potentials leveraging the GAP methodology have been developed based on SOAP descriptors
employing values of $\zeta=2$ or $\zeta=4$.

Finally, let us summarise this whole procedure of obtaining
the SOAP descriptor which are also outlined in the bottom of figure \ref{fig:acsf-soap}:
\begin{enumerate}
    \item Obtain element-specific neighbour densities for each atom in the system.
    \item Expand these densities in a radial and angular basis and obtain the resulting expansion coefficients.
    \item Build the so-called power spectrum, by combining the expansion coefficient of different elements and radial channels while averaging over the angular components.
    \item Normalise the resulting SOAP vector to enable direct comparison between different atomic environments.
\end{enumerate}
While being conceptually a bit more involved than ACSFs, they
have the advantage that they are more systematic in the sense
that they rely only on two hyperparameters $n_\text{max}$ and $l_\text{max}$ (assuming we have a physically motivated guess for $\sigma_a$)
controlling the radial and angular expansion.
\subsection{Discussion and Outlook on Chemical Descriptors}
\label{sec:desc_outlook}
We have seen that the introduction of a set of general body-ordered
functions to transform Cartesian coordinates into an atomic fingerprint
enables the required invariances for the representation
of PESs with data-driven approaches to be included.
From a first glimpse, this might seem quite elaborate as we are
transforming $3N_\text{atom}$ coordinates into $N_\text{atom}\cdot N_\text{des}$ descriptors, where the length of the descriptor can easily
surpass the thousands.
This means that we are arriving at an overcomplete description of our system.
However, this additional sensitivity is actually quite useful as it means our
regression task can become much easier than for the minimal number of $3N_\text{atom}-6$ internal degrees of freedom.
Moreover, it allows the application of the MLP to systems of -- in principle -- arbitrary size rather than being constrained to the exact number of atoms encountered in systems of the training set.

It is important to note that the transformation via descriptors is not for free.
In many cases, it is actually the most costly part of the evaluation of an MLP.
In general, as higher body terms are included in the descriptors, the more expensive the computation of the descriptors becomes, due to the steeply rising combinatorial complexity.
However, there are certain tricks that can dramatically reduce the cost, which we will come back to in Sec~\ref{sec:devel}.

While being very successful and widely used in many applications,
the two presented descriptors have some limitations.
One shortcoming of both ACSFs and SOAP vectors is the unfavourable
scaling with the number of species in the system.
This means that systems on the order of 5-6 different chemical species are currently
the limit of what has been described with these descriptors.
One way to solve this problem is to account for the element-specific composition
of the chemical environment in an implicit manner 
by introducing element-dependent weighting functions in the descriptors,
instead of using separate functions to describe different
combinations of elements, as done in weighted ACSFs~\cite{Gastegger2018/10.1063/1.5019667}.
In addition, there are new ideas to sparsify the descriptors and
thus break the unfavourable scaling, such as done in
tensor-reduced atomic density representations~\cite{Darby2022/10.48550/arxiv.2210.01705}.
Furthermore, both types of descriptors discussed above have recently been
shown to be incomplete in the sense that they can have trouble differentiating
certain arrangements of atoms.~\cite{Pozdnyakov2020/10.1103/PHYSREVLETT.125.166001}
A solution to this is given by descriptors used in moment tensor potentials~\cite{Shapeev2015/10.1137/15M1054183},
or the more recent atomic cluster expansion (ACE)~\cite{Drautz2019/10.1103/PHYSREVB.99.014104}, which both include
in principle, higher-order body terms up to infinite order.
In practice, they are still truncated after a certain order, nevertheless
making them more complete and adjustable to the system of interest.
The ACE approach, which is described in more detail in section~\ref{subsec:ace},
has the additional advantage of allowing for equivariance of the output
properties with respect to the coordinates of the system.
This enables direct representation of vectorial and tensorial properties,
such as dipole moments or polarizabilities, which rotate with the system
of interest.

An additional shortcoming in the most literal sense is the restriction to
a certain cutoff around each atom, making the resulting descriptors
short-sighted.
This locality approximation is known to be problematic for cases
where long-range effects such as Coulomb interactions are important
for the system of interest.
They can be included by a suitable baseline, for example by learning
partial charges given by reference 
calculations~\cite{Artrith2011/10.1103/PhysRevB.83.153101,
Ko2021/10.1038/s41467-020-20427-2}, or special long-ranged
descriptors such as the long-distance equivariant (LODE) framework~\cite{Grisafi2019/10.1063/1.5128375}.
Other developments make the descriptors a part of the ML model
thus learning them as part of the training process of the model,
for example in the context of message-passing neural networks (MPNNs)~\cite{Gilmer2017}. 
Popular examples of these MPNNs include SchNet \cite{Schutt2018/10.1063/1.5019779}, PhysNet \cite{Unke2019/10.1021/ACS.JCTC.9B00181}, and NequiP \cite{Batzner2021/10.1038/s41467-022-29939-5}.
While SchNet and PhysNet provide invariant representations of the atomic structure, NequiP introduces a significant innovation by directly incorporating equivariant features that embed rotational symmetry into the model.
Furthermore, the recently introduced Allegro model~\cite{musaelian_learning_2023} learns the atomic representation without relying on atom-centred message passing, offering an efficient alternative.
We will come back to these ideas in section~\ref{subsec:mess} in more detail.

\section{Regression Models}
\label{sec:regression}
After having seen how chemical structure can be encoded based on descriptors,
we will next focus on the actual machine learning approaches to relate
structural information to the PES.
As we are after a description of a continuous property, i.e. energy, we are dealing with
a regression task.
The simplest way to achieve this would be by linear regression,
where the energy $E$ is obtained through a linear relation of the descriptor values $G$
\begin{align}
    E = \sum_{i}^{N_\text{atom}} \sum_{j}^{N_\text{des}} G^{j}_{i} \cdot a^{j}_{i} (+ b^{j}_{i}),
\end{align}
summing over all $N_\text{atoms}$ atoms in the system and all $N_\text{des}$ associated descriptor values of these atoms.
Here, $a$ and $b$ are the only parameters of the model optimised to reproduce the reference energies.
In cases where the descriptors are very high-dimensional and sensitive,
this can actually give very good representations of the PES after one important modification.
If the parameters $a$ and $b$ of the model are chosen to be different for each atom,
we are loosing the permutational invariance that we painstakingly introduced
in the previous section.
To illustrate that let us consider a simple system of two atoms:
In this case, the order of our descriptor vectors $G_i$ can't be swapped
after the weights have been optimised.
Most MLPs do not rely on relatively restricted linear regression and
instead apply more flexible regression models, i.e. models with orders of magnitude more parameters to optimise.
Out of these, the most commonly used approaches in the field fall into two broad categories:
Artificial neural networks, or kernel-based methods.
In the following, both of these regression models will be introduced
and we will see how they can be used in MLPs
to represent PESs.

\subsection{Artificial neural networks}
\label{sec:ann}

Artificial neural networks, as one of the prominent models in machine learning,
originated as a representation of biological neural networks,
but are nowadays a widespread machine learning model capable
of reproducing highly complex relations between an input and an output without
knowledge about the underlying causation\cite{Haykin2008}.
Their functional form is based on a layered structure
with connected nodes and associated weights as schematically depicted
in Fig.~\ref{fig:ann_fit}.
The nodes of the input layer, which hold the provided mathematical
description of the input (here: structural information of atoms encoded in descriptor space),
are connected via so-called
hidden layers to the nodes of the output layer, which after evaluation
of the associated functional form contain the information (here: potential energy) that
should be associated with the input.
In feed-forward neural networks, which are the most used form for representing PESs,
all node values in a layer depend exclusively on the nodes of the preceding
layer~\cite{Haykin2008}.
All of the associated connections hold a corresponding weight~$a$.
Additional bias nodes are usually connected to all nodes except the input
and serve as an adjustable offset to shift the input of the
individual nodes via bias~weights~$b$.
The functional relation for the evaluation of a node value
is essentially a linear dependence on all node values of the previous layer
that is further ``activated'' by a so-called (typically non--linear)
activation function as shown on the right side of Fig.~\ref{fig:ann_fit}.
The value $y^{i}_{j}$ of node $j$ in layer $i$ is then calculated as
\begin{align}\label{eq:node}
    y^{i}_{j} = f^{i}_{j}\left(b^{i}_{j}+ \sum_{k=1}^{n_{i-1}} a^{i-1,i}_{k,j} \cdot y^{i-1}_{k}  \right) 
\end{align}
implying that only the bias as well as the respective node values
and weights of the previous layer contribute to the values of the layer of interest.
Here, $k$ is then the index over the nodes in the preceding layer that features $n_{i-1}$ nodes.

The function $f^{j}_{i}$ is usually a sigmoid function in
the hidden layers, which can be represented as a hyperbolic tangent,
and the linear function $f(x)=x$ for the output layer to
allow for continuous predictions.
The characteristics of a sigmoid function is that it switches
in a small interval from 0 to 1 (or similar) and stays constant
outside of that interval.
This function is responsible for the great flexibility of neural networks.
Without the activation function, the model would feature only highly
convoluted linear dependencies with rather poor capability to represent
other non-linear functional forms.
Due to the activation by the sigmoid function of the gathered linear
information from previous layers for each node, neural networks
are able to fit arbitrary functional forms
and are therefore ideally suited for machine learning.
Note that there are also other choices of activation functions
such as the rectified linear unit (ReLU) $f(x)=max(0,x)$,
or related variants.
In earlier days, these functions were not commonly used as they are
not differentiable at all points.
However, in recent times they have become a very popular choice
in particular, in deep learning applications, given their
speed and suitability for graphical processing units.

Predictions with a neural network happen via forward-passing
of the input information via the different layers in the neural network.
It is important to note, that this functional form
can be evaluated very efficiently by
the usage of a vector matrix representation of the
nodes and weights.
Similarly, derivatives of the output can be evaluated
efficiently via so-called backpropagation where 
the partial derivatives are evaluated layer by layer
and passed backwards through the network.
This enables efficient calculation of derivatives
with respect to the network parameters, required 
for an optimisation of the network, but also with
respect to the structural information required
for obtaining forces of a PES.
Backpropagation is nowadays a standard feature of
modern machine learning libraries such as Pytorch or
Tensorflow and can be executed with a simple functional call.
For further details on neural networks, the reader is referred
to Ref.~\citenum{Haykin2008}, while the
application of this model to chemical systems will be
presented in the following.

\subsubsection{High--dimensional Neural Network Potentials}
\label{sssec:hdnnp}
Let us next see how neural networks can be used to represent the
PES of a system of interest.
In principle, the structural information of each atom is encoded by
the descriptors of choice could be used directly as input
of a global neural network that would output the potential energy.
Such an approach was actually used in early studies of representing
PESs with machine learning techniques.~\cite{Blank1994/10.1002/cem.1180080605,Witkoskie2005/10.1021/ct049976i}
However, this results in complications,
as the order of the atoms will matter in such a setup.
This means that invariance with respect to permutations of atoms
is again lost.
Furthermore, the complexity of the required model scales very
unfavourably with the number of atoms in the system, making this approach unfeasible for larger systems.
Behler and Parrinello managed to incorporate the required
structural invariances and break the unfavourable scaling
by representing every element of the system
by a separate NN resulting in the so-called high--dimensional neural network
potential approach~\cite{Behler2007/10.1103/PhysRevLett.98.146401,
Behler2021/10.1021/acs.chemrev.0c00868}.
For this purpose, the total potential energy of a particular configuration is separated
into the sum of the contributions of individual atoms to
construct the functional relation between
the energy as output and atomic configuration as input
\begin{align} \label{eq:esum}
    E_\text{tot} = \sum_{i=1}^{N_\text{atom}}E_{i} = \sum_{s=1}^{N_\text{ele}} \sum_{i_s=1}^{N_{\text{atom}_s}} E_\text{NN$_s$}(G_{i_s}) .
\end{align}
Each atomic contribution for a particular element $s$ is represented with a single neural network ($E_\text{NN$_s$}$), shared for all atoms of that element ($N_{\text{atom}_s}$).
This also enables systems of different sizes to be represented with
the same model, or to apply the trained model to
a larger system, as long as the required chemical environments
are covered in the training set.
In addition, it means that there is linear scaling with
the number of atoms in the system, as every atomic contribution
can be evaluated independently from the others.
A schematic depiction of the resulting representation
is shown in the upper part of Fig.~\ref{fig:hdnnp-gap} for the example
of a water slab.

Representing each element with a different NN not only
introduced the permutational invariance, but is also an
intuitive division of the potential energy as atoms of the
same elements are usually more similar than of other elements.
However, it is important to note that the resulting atomic energies
are not physical as they are simply a tool to reproduce the
correct total potential energy.
The partitioning is merely a requirement by the locality approximation and a consequence of the fitting
and different starting initialisation will result in different
atomic energy contributions,
as for example shown in Fig. 12.4 of Ref.~\citenum{Gastegger2020/10.1007/978-3-030-40245-7_12}.
Nevertheless, there are reports in the literature
that showcase the usefulness of atomic energies as an
analytical tool and analyse the robustness of local predictions~\cite{Bernstein2019/10.1002/ANIE.201902625,Chong2023}.

This high-dimensional NN scheme is local in that the energy of an atom depends only on the atoms in the close
neighbourhood.
This is a necessary condition to reduce the effective
the dimensionality of the problem, which is intractable
otherwise.
It is important to note that such a reduction is done
in all types of empirical potentials in some way.
However, an immediate consequence
is the lack of any long-range interactions, such as
electrostatics.
Let us summarise this approach once more:
A set of descriptors for each element transforms
the coordinates of the system to be employed as input for the HD-NNP.
These vectors serve as the input for the atomic NNs,
which yields the atomic energy contributions
that sum up to the total observable as schematically
depicted in the upper part of Fig.~\ref{fig:hdnnp-gap}.
This functional form additionally allows analytical gradients
to be calculated, which is an important feature for
molecular dynamics-based sampling techniques.
In practice, the values of each symmetry function are usually
centred around the average value of the training set and
normalised to values between zero and one.
This has the advantage that all symmetry functions are on a similar scale
and their range falls naturally within the steep
region of the activation function, making it easier to find
optimal parameters and preventing ``saturation'' of the network,
meaning that most of the hidden nodes have values close to -1.0 or +1.0.
Usual architectures for the atomic neural networks
in the HD-NNP formalism
feature two to three hidden layers with 25 to 40 nodes each.
These might seem small compared to other applications of neural networks,
but is sufficient in this case due to the choice of suitable descriptors.
Nevertheless, other more recent methods feature much larger
and deeper neural networks, depending on the system of interest
and choice of descriptors.
Bias nodes with weight parameters $b$ are also commonly attached
to all nodes except those in the input layer.
Furthermore, activation functions are not applied in the output
layer as this would strongly limit the possible predictive range
making it impossible to learn arbitrary potential energy surfaces.
\subsubsection{Obtaining Analytical Derivatives}
In atomistic simulations, forces are generally obtained by computing the negative gradient of the total energy with respect to atomic positions $F_i = -\nabla_i E$.
To obtain the force acting on atom $i$ with respect to some Cartesian coordinate $\alpha$ we must apply the chain rule since we have transformed from Cartesian coordinates into symmetry function representations;
\begin{align}
    F_{i,\alpha} &= -\frac{\partial E}{\partial \alpha} = -\sum_{i=1}^{N_\text{atom}}\frac{\partial E_i}{\partial \alpha} \\ \nonumber
    &= -\sum_{i=1}^{N_{atom}}\sum_{s=1}^{N_{des,i}}\frac{\partial E_{NN_i}}{\partial G_{i,s}}\cdot \frac{\partial G_{i,s}}{\partial \alpha}
\end{align}
using the fact that the total energy $E$ is a sum of atomic contributions $E_i$ from Equation \ref{eq:esum} and $G_{i,s}$ is the $s^{th}$ symmetry function of the $N_{des,i}$ symmetry functions for a given atom $i$.\\

The first component $\partial E_{NN_i}/\partial G_{i_s}$
depends on the architecture of the atomic neural networks and is obtained from the backpropagation of the partial derivatives through the network.
\rev{As mentioned above, this can nowadays be efficiently obtained
by directly calling a `backprop' function offered by modern machine learning
libraries such as Pytorch or Tensorflow.}
Rewriting Equation \ref{eq:node} for a given layer $n$ we get 
\begin{align}
     y^n = f\left(z_{n-1} \left(f\left( z_{n-2}...\left(f\left(z_1\left(G_{i,s}\right)\right)\right)\right)\right)\right)
\end{align}
where in matrix notation
\begin{equation}
    z_i = (\mathbf{b}_i+ \mathbf{a}_{i}^{\mathrm{T}}\cdot \mathbf{y}_{i-1}).
\end{equation}
The partial derivative of the NN output node with respect to the symmetry functions is then
\begin{align}
    \frac{\partial y^n}{\partial G} = \frac{\partial f}{\partial z_{n-1}}\frac{\partial z_{n-1} }{\partial f} ... \frac{\partial f}{\partial z_1}\frac{\partial{z_1}}{\partial G}.
\end{align}

The second term $\partial G_{i,s}/\partial \alpha$
depends on the various radial and angular symmetry functions describing the atom.
Considering for example the radial component of the ACSF given in Equation \ref{radsymf} the derivative on the central atom $i$ with respect 
to some coordinate $\alpha$ which can be any of $x$,$y$,$z$ component of atoms $i$ or $j$
\begin{align}
    \frac{\partial G^{\mathrm{rad}}_i}{\partial \alpha} = & \sum_j \bigg[2\eta (R_{ij}-R_s) f_{c}R_{ij}e^{-\eta(R_{ij}-R_s)^2}\cdot \frac{\partial R_{ij}}{\partial \alpha} \\ \nonumber
     & + e^{-\eta(R_{ij}-R_s)^2}\cdot \frac{\partial f_c(R_{ij})}{\partial \alpha}\bigg].
\end{align}
It is easy to show from the definition of $R_{ii}$
\begin{equation}
    R_{ij} = \sqrt{(x_i-x_j)^2 + (y_i-y_j)^2 + (z_i-z_j)^2}
\end{equation}
that
\begin{align}
    \frac{\partial{R_{ij}}}{\partial x_i} & = \frac{1}{R_{ij}}(x_i - x_j)
\end{align}
and
\begin{equation}
    \frac{\partial R_{ij}}{\partial x_i} = - \frac{\partial R_{ij}}{\partial x_j}.
\end{equation}
Computing $\partial G^{\mathrm{ang}}_i /\partial \alpha$ is similarly straightforward however more algebraically involved, and so the reader is referred to the SI of Ref.~\citenum{Singraber2019/10.1021/ACS.JCTC.8B00770}.
Similarly, the stress tensor, important for the pressure response
of a system, can be evaluated analytically via
suitable partial differentiation as described in detail in
Ref.\citenum{Behler2011/10.1063/1.3553717}
purely based on pair contributions.
\subsubsection{Training}
\label{sec:training}

After setting up a particular HD-NNP architecture,
the model needs to be ``trained'' to reference points
by optimising its parameters in an iterative
procedure to reproduce the correct total energies (and forces) of the training set.
This is done by minimising a loss function, $L$, defining the 
fitness of a model to reproduce the given reference values:
\begin{align}
    L = \frac{1}{2N} \sum_{i=1}^{N} \left(E^\text{ref}_{i} - E^\text{model}_{i}\right)^2.
\end{align}
Such loss functions can also be augmented by forces, where
there is then some freedom about weighting the two types of
information.
Targeting forces together with energies during the optimization
of an MLP usually leads to much better fits as the forces
provide additional information about the curvature of the PES.
Furthermore, they are local properties associated with each atom
which increases the information content of a single configuration
vastly.
\rev{Training on both energies and forces while retaining energy conservation is possible due to the analytical link between the two properties as shown in the previous section.}

Once equipped with a suitable loss function, the weights of the model need to be updated in order to minimise the loss.
A simple way to achieve this is local optimisation techniques like steepest descent, where the gradient of the loss with respect to the weights $w$ is followed towards a lower loss
\begin{align}
    w_\text{new} = w_\text{old} - \eta \frac{\partial L}{\partial w_\text{old}}, 
\end{align}
where $\eta$ is an adjustable learning rate.
Since the loss function is a simple sum over all instances in the training set,
these updates can be performed consecutively for each structure.
The optimisation process is then usually grouped into so-called epochs,
where one epoch means that all structures in the training set were considered once to update a given model.
The number of epochs is then another hyperparameter that needs to be chosen
by the user.

\begin{figure*}[tb]
    \centering{}
    \includegraphics[width=1.0\textwidth]{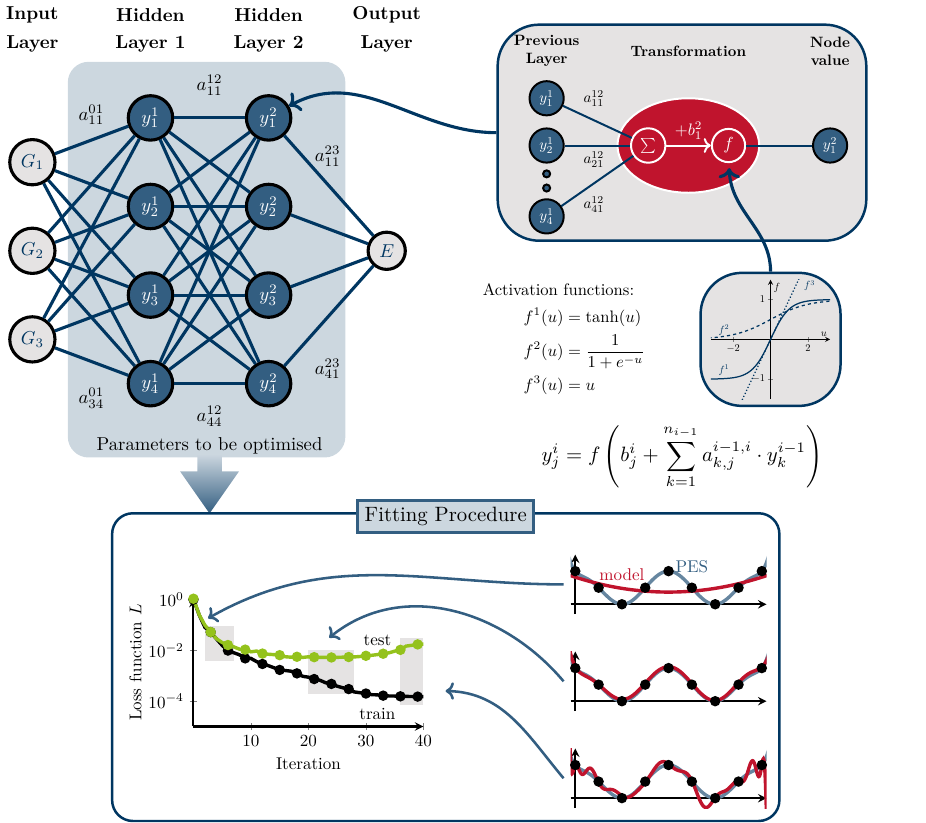}
    \caption{\textbf{Schematic of typical neural network architecture (top) and its typical optimization process (bottom).}
        Top left: Schematic representation of a neural network with two hidden
        layers. The output $E$ is obtained as a function of the
        three input nodes $G$ with four nodes $y$ in each hidden layer.
        The weights $a$ are represented by lines, while bias weights
        $b$ are omitted for clarity.
        Top right: Illustration of the functional dependence of a single
        node output value on the node values of the previous layer.
        Bottom left: Illustration of a typical optimization process for a neural network
        model. Left: Mean error of the training (black)
        and test set (green) in the progress of the optimization.
        After substantial improvement of both errors at the beginning
        of the optimization, the training error usually keeps
        decreasing, while the test error stagnates or even increases
        after some time.
        Bottom right: Corresponding comparison of the reference function (blue)
        to the neural network model (red) during three representative
        stages of the optimisation.
        The neural network is trained to a set of training points
        of the reference function (black dots) for which the prediction
        is continuously improved.
        Due to the high flexibility of the model, regions
        in between the training points deteriorate after too heavy
        optimisation.
    }
    \label{fig:ann_fit}
\end{figure*}

Usually, a set of test structures, different from the training set,
is prepared as well to estimate the quality of the fit.
This test set provides an estimate for the transferability
to structures not included in the training set.
Due to the high flexibility of the model, they
can represent rather complex functional relations, but are
also prone to overfitting~\cite{Hawkins2004/10.1021/CI0342472}
and have usually many
local minima in parameter space.
Simple optimiser like gradient-descent-based methods
will therefore usually worsen the accuracy of the 
prediction of the test set after some time.
To illustrate this, the typical performance of an ML model
during optimisation is shown at the bottom of Fig.~\ref{fig:ann_fit}.
While the training points are reproduced better and better
in the progress of the optimisation, the high flexibility
of the model leads to large fluctuations in between the
training points, which can be detected by an increase
of the error in the test set.

One method to circumvent these problems is to stop early during
the optimisation and carefully monitor the test error
for signs of overfitting.
In addition, more efficient optimiser
such as the adaptive global extended Kalman
filter~\cite{Shah1992/10.1016/S0893-6080(05)80139-X,Singraber2019/10.1021/ACS.JCTC.8B00770}
significantly improve the performance of the fitting procedure and help to
prevent overfitting from points far away from a given training instance.
However, this does not change the fact that many
quasi--degenerate local minima are located
in the parameter space of the model.
It is therefore good practice to perform a
variety of fits with different starting conditions
to select the set of parameters that is optimal
for the training and test set, although this
will most certainly still not be the global minimum
of the parameters.~\cite{Behler2017/10.1002/ANIE.201703114}
Since the model only provides a mathematical representation
of the envisaged relation and can not be used to
infer causation, this does not have further
consequences.

In contrast, for kernel-based approaches, introduced in the next section,
the training procedure is usually more straightforward.
In most cases, there exists a closed-form analytical solution
which can be directly evaluated to get the global minimum
without the need for iterative local optimisation.

\subsection{Gaussian Process and Kernel Ridge Regression}
\label{sec:gpr}

Gaussian process regression (GPR) represents a Bayesian nonparametric regression
technique able to approximate complex nonlinear functions of high dimensionality.~\cite{Rasmussen2018/10.7551/mitpress/3206.001.0001}
GPR is similar to kernel ridge regression (KRR) in the sense that they
both provide the same type of predictions.
However, GPR goes a step further by also giving an estimate of the uncertainty
of the prediction, which KRR lacks.
Generally, the GPR framework can be derived based on two different approaches,
the so-called \textit{weight}-space and \textit{function}-space views. 
Following closely the excellent review by Deringer and coworkers~\cite{Deringer2021}, here we highlight elements from both routes to best communicate the conceptual idea
behind using GPR to \textit{learn} the PES of a system of interest based on 
quantum mechanical reference data.

\subsubsection{Weight-Space View}
\label{sec:weight-space}
In the weight-space view on GPR, we start by approximating the high-dimensional
function $y(\textbf{x})$ as a linear combination of $M$ basis functions, $k$,
centered at representative locations $\textbf{x}_m$ of the input space, such as:
\begin{equation}
    y(\textbf{x}) \approx \rev{f}(\textbf{x}) = \sum_{m=1}^M c_m k(\textbf{x}, \textbf{x}_m) \mbox{ , }
    \label{eq:GPR_weight_basis}
\end{equation}
where these so-called kernel functions $k$ measure the similarity between
two arbitrary data points and $c_m$ are the corresponding coefficients.
While the functional form of $k$ does not matter for this derivation, for the sake
of illustration, we can imagine Gaussians being placed at the set of representative
locations, $\{\textbf{x}_m\}_{m=1}^M$.
This is commonly referred to as the Gaussian or square exponential kernel. 
It is important to stress, however, that $k$ needs to be symmetric and
positive semi-definite.
The coefficients, $c_m$, are then obtained by fitting 
equation \ref{eq:GPR_weight_basis} to a training set,
$ \mathcal{D} = \{\textbf{x}_n; y_n\}_{n=1}^N$, which
corresponds to minimising the loss function
\begin{equation}
    \mathcal{L} = \sum_{n=1}^N \frac{\left[ y_n - \rev{f}(\textbf{x}_n)  \right]^2}{\sigma_n} + \sum_{m,m'}^M c_m k(\textbf{x}_m, \textbf{x}_{m'}) c_{m'} \mbox{,}
    \label{eq:GPR_weight_loss}
\end{equation}
where the first term aims to minimise the difference between the data
and GPR model and the second term is a (Tikhonov) regularisation
to ensure small coefficients, $c_m$, and prevent overfitting.
The effect of the regularisation can be best understood by remembering
that the kernel function $k$ measures the similarity between two points $m$ and $m'$.
If $m=m'$, the value is one and the squared coefficient contributes
strongly to the loss.
Minimisation of the loss thus results in keeping $c_m$ small.
If we have very dissimilar points (small $k$), their coefficient would need
to be rather large to provide a meaningful contribution to the
prediction.
However, also this is disfavoured by the regularisation term,
thus preventing overfitting.
The parameter $\sigma_n$ weights the importance of the $n$th data point
and implicitly determines the strength of the regularisation term;
choosing $\sigma_n$ carefully is important to obtain an accurate
and smooth GPR model.
The result of this is that we effectively interpolate for
an unknown point $\mathbf{x}_\text{new}$ using a linear combination
of the most similar representative locations $\mathbf{x}_m$, given
that small similarities will require large coefficients which
are suppressed by the regularisation.

To facilitate the next steps, we rewrite equation \ref{eq:GPR_weight_loss}
in matrix notation
\begin{equation}
    \mathcal{L} = \left[ \textbf{y} - \textbf{K}_{NM} \textbf{c} \right]^T \bm{\Sigma} \left[ \textbf{y} - \textbf{K}_{NM} \textbf{c} \right] + \textbf{c}^T \textbf{K}_{MM} \textbf{c}  \mbox{,}
    \label{eq:GPR_weight_loss_matrix}
\end{equation}
where the $\bm{\Sigma}$ is a diagonal matrix of size $N$ with the diagonal elements
$\bm{\Sigma}_{nn}=\sigma_n^2$.
The elements of the kernel matrix represent the similarity between
each pair of input locations, $\textbf{x}_n$ and $\textbf{x}_m$,
as determined by the kernel function, $k(\textbf{x}_n, \textbf{x}_m)$. 
Furthermore, the matrix is symmetric, resulting in 
$\textbf{K}_{NM}^T = \textbf{K}_{MN}$.
The coefficients can then be obtained by setting the  derivative
of the loss function with respect to the coefficients to zero.
These are given by
\begin{equation}
    \textbf{c} = \left[\textbf{K}_{MM} + \textbf{K}_{MN} \bm{\Sigma}^{-1} \textbf{K}_{NM} \right]^{-1} \textbf{K}_{MN} \bm{\Sigma}^{-1} \textbf{y} \mbox{,}
    \label{eq:GPR_weight_coeffs}
\end{equation}
which can be used to make predictions using the matrix notation
of equation \ref{eq:GPR_weight_basis}
\begin{equation}
    \rev{f} (\textbf{x}) = \textbf{c}^T\textbf{k}\mbox{ , }
    \label{eq:GPR_weight_basis_matrix}
\end{equation}
with $\textbf{k}$ being the \rev{shorthand notation for the} vector of kernel values for
$\textbf{x}$ with respect to the representative points, i.e.
$[\textbf{k}(\textbf{x})]_m = k(\textbf{x}, \textbf{x}_m)$. 
Unlike artificial neural networks, where weights are typically
found through numerical optimisation that usually leads to local minima, 
the coefficients in equation \ref{eq:GPR_weight_coeffs} are derived 
analytically and represent the global optimum based on the input data,
the chosen kernel functions and representative points, and the hyperparameters
(both within the kernel function and $\sigma_n$).
Setting the number of basis functions, $M$, appropriately is important in GPR. 
It may seem tempting to place a basis function at every data input location
for accurate representation of $y(\textbf{x})$.
With this strategy ($M=N$), known as full GPR, the expression for the coefficients in equation \ref{eq:GPR_weight_coeffs} simplifies to 
\begin{equation}
    \textbf{c} = \left[\textbf{K}_{NN} + \bm{\Sigma}\right]^{-1}  \textbf{y} \mbox{,}
    \label{eq:gpr_weight_training_coeff_full_gpr}
\end{equation}
and the GPR prediction at location $\textbf{x}$ is given by 
\begin{equation}
    f(\textbf{x}) = \textbf{k}^T\left[\textbf{K}_{NN} + \bm{\Sigma}\right]^{-1}  \textbf{y} \mbox{.}
    \label{eq:GPR_weight_pred_fullGPR}
\end{equation}
Full GPR, however, is impractical for large data sets due to the high computational cost and memory demands associated with inverting the matrix $\textbf{K}_{NN}$ during training (equation \ref{eq:gpr_weight_training_coeff_full_gpr}), scaling as $\mathcal{O}(N^3)$ and $\mathcal{O}(N^2)$, respectively.
Once the coefficients $\mathbf{c}$ are fixed after fitting, the inference time for making predictions (equation \ref{eq:GPR_weight_basis_matrix}) is determined by the computation of the vector $\mathbf{k}$ and scales linearly with the number of representative points used, which is $\mathcal{O}(N)$ for full GPR.
Therefore, most applications opt for sparse GPR (i.e. $M<<N$) where
the number and location of the basis functions are chosen
strategically to balance between accuracy and computational efficiency.
\rev{In addition to faster inference, sparse GPR also facilitates fitting larger data sets, as the computational costs associated with determining the coefficients (as in Equation \ref{eq:GPR_weight_coeffs}) scale as $\mathcal{O}(NM^2)$.}
By now, we have reached the point where KRR and GPR diverge;
Although the predictions produced by combining equations \ref{eq:GPR_weight_coeffs}
and \ref{eq:GPR_weight_basis_matrix} are the same for both KRR and GPR, we will
make a short excursion on the function-view on GPR to comprehend how it provides
an uncertainty estimate.

\subsubsection{Function-Space View}
\label{sec:func-space}
In this section, we depart slightly from the review of Deringer et al.~\cite{Deringer2021} and present an alternative derivation for the function space view inspired by the comprehensive textbook of Rasmussen and Williams \cite{Rasmussen2018/10.7551/mitpress/3206.001.0001}. 
However, we recommend exploring both derivations for a complete grasp of the involved concepts.
In the function-space view, we shift from regarding the estimator $f(\textbf{x})$ as a fixed deterministic mapping to a probabilistic description through a Gaussian process (GP)
\begin{equation}
    y(\textbf{x}) \approx f(\mathbf{x}) \sim \mathcal{GP}(\mu(\mathbf{x}), \mathrm{Cov}[f(\mathbf{x}), f(
\mathbf{x'})] )\mbox{ . }
    \label{eq:GPR_gaussian_process}
\end{equation}
A GP is a distribution over functions consistent with data, where any finite set of function values is drawn from a joint (multivariate) Gaussian distribution.
This allows us to capture not just a single function but an entire ensemble of possible functions that align with the available data.
Before conditioning on data, the GP is fully defined by its mean function, $\mu(\mathbf{x})$, and covariance function, $\mathrm{Cov}[f(\mathbf{x}), f(\mathbf{x'})]$.
For the sake of clarity, here we take the mean function to be zero, however, if there is a good guess available, the mean can be subtracted from the observed function values before fitting and 
added back after prediction.
The covariance, conversely, is described by a kernel function, $k$, which quantifies the similarity between function values at different points:
\begin{equation}
    \mathrm{Cov}[f(\mathbf{x}), f(
\mathbf{x'})] = k(\mathbf{x},\mathbf{x'}) \mbox{ , }
    \label{eq:GPR_func_kernel}
\end{equation}
ensuring the smoothness of the function $f$ across the input space.
The resulting multivariate Gaussian distribution is called the (GP) prior and incorporates our initial assumptions about the functions we are dealing with \textit{before} observing any data.
Thus, any admissible function computed at arbitrary locations $\{\textbf{x}_m\}_{m=1}^M$ is given by the joint Gaussian distribution:
\begin{equation}
   \mathbf{f} = [f(\textbf{x}_1), \ldots, f(\textbf{x}_N)] \sim \mathcal{N}(\mathbf{0}, \textbf{K}_{MM}) \mbox{ , }
   \label{eq:GPR_func_joint_example}
\end{equation}
where $\textbf{f}$ is the vector of function values and $\textbf{K}_{MM}$ is a kernel matrix as introduced in the previous section.\\

Rather than drawing random functions from the prior, we are usually interested in incorporating knowledge from the observations in our training set, $\mathcal{D} = \{\textbf{x}_n; y_n\}_{n=1}^N$, to make predictions at an unseen location, $\textbf{x}_{\star}$.
In other words, we are after the conditional probability distribution $P(f(\textbf{x}_{\star})| \textbf{y})$ which is also  Gaussian and can be easily constructed from the joint GP prior distribution of the training outputs, $\textbf{y}$, and the predicted output at the new location, $f(\textbf{x}_{\star})$.
In this context, it is important to note that the observed outputs are often considered as noisy versions of the true underlying function values, such that $\textbf{y}=f(\textbf{x})+\epsilon$, where $\epsilon$ is independent identically distributed Gaussian noise with zero mean $\sigma^2$ variance.
Then, the covariance function of two measurements yields
\begin{equation}
\mathrm{Cov} \left[y_n, y_{n'}\right] = k (\textbf{x}_n, \textbf{x}_{n'}) + \sigma^2 \delta_{nn'} \mbox{ , }
\label{eq:GPR_funct_cov_noise}
\end{equation}
which we can rewrite in matrix notation as $\textbf{K}_{NN}+ \sigma^2 \textbf{I}$. 
Similar to the weight-spaced view, we can employ data-point-specific variances for
the noise, $\sigma_n$, allowing us to replace $\sigma^2 \textbf{I}$
with $\bm{\Sigma}$.\\

\rev{Now, the joint distribution of the noisy training outputs, $\textbf{y}$, and the prediction, $f(\textbf{x}_{\star})$ according to the prior is
\begin{equation}
\begin{bmatrix}
\mathbf{y} \\
f(\mathbf{x}_{\star})
\end{bmatrix}
\sim \mathcal{N}\left(
\mathbf{0},
\begin{bmatrix}
\textbf{K}_{NN} + \bm{\Sigma} & \mathbf{k} \\
\mathbf{k}^T & k(\mathbf{x}_{\star},\mathbf{x}_{\star})
\end{bmatrix}
\right) \mbox{ , }
\label{eq:GPR_func_joint}
\end{equation}
where $\textbf{k}$ is the shorthand notation for the vector of kernel values for $\mathbf{x}_{\star}$ with respect to the locations encountered in the training data, similar to the definition in the previous section.
Then, the conditional distribution of $f(\mathbf{x}_{\star})$ given the training outputs, $\textbf{y}$, is
\begin{align}
    f(\mathbf{x}_{\star}) | \textbf{y} &\sim \mathcal{N}\left( \bar{f}(\mathbf{x}_{\star}), \mathrm{var}[f(\mathbf{x}_{\star})]\right) \mbox{ , } 
\end{align}
with
\begin{align}
    \bar{f}(\mathbf{x}_{\star}) &=\textbf{k}^T [\textbf{K}_{NN} + \bm{\Sigma}]^{-1} \textbf{y} \mbox{ , } \label{eq:GPR_funct_mean}\\
    \mathrm{var}[f(\mathbf{x}_{\star})]  &= k(\mathbf{x}_{\star},\mathbf{x}_{\star}) - \textbf{k}^T [\textbf{K}_{NN} + \bm{\Sigma}]^{-1} \textbf{k} \mbox{,} \label{eq:GPR_funct_var}
\end{align}
which correspond to the key predictive equations in GPR which is commonly referred to as predictive distribution. A detailed derivation of conditioning the joint GP distribution on the observations can be found in \cite{Rasmussen2018/10.7551/mitpress/3206.001.0001}.
Our best estimate of $f(\mathbf{x}_{\star})$ is then the mean of this distribution, given in equation \ref{eq:GPR_funct_mean}, and associated uncertainty is quantified by the variance in equation \ref{eq:GPR_funct_var}.}
As expected, this prediction of the GPR model is equivalent to that obtained in the weight\rev{-}space view when full GPR ($M=N$) is employed given by equation \ref{eq:GPR_weight_pred_fullGPR}.
Interestingly, the variance only depends on the location and similarity of the data points and the noise associated with each point, but not the training
set values.
This provides us, in principle, with the necessary tools to perform GPR.
However, there is one important aspect related to fitting to quantum mechanical
reference data which needs addressing.
As seen for HD-NNPs, machine learning potentials are best setup
to estimate an atomic energy function given an
atom's local environment, but they are constructed based on electronic structure
calculations yielding only the \textit{total} energy and its derivatives, 
namely the atomic forces and virial stresses. 
Therefore, 
some more steps are needed to be able to learn
a function when the function's actual values are not
available to us, but we have access to its derived properties.
We refer the interested reader to the excellent review by Deringer and coworkers
for further details~\citenum{Deringer2021}.

\begin{figure*}[t]
\centering
\includegraphics[width=1.0\textwidth]{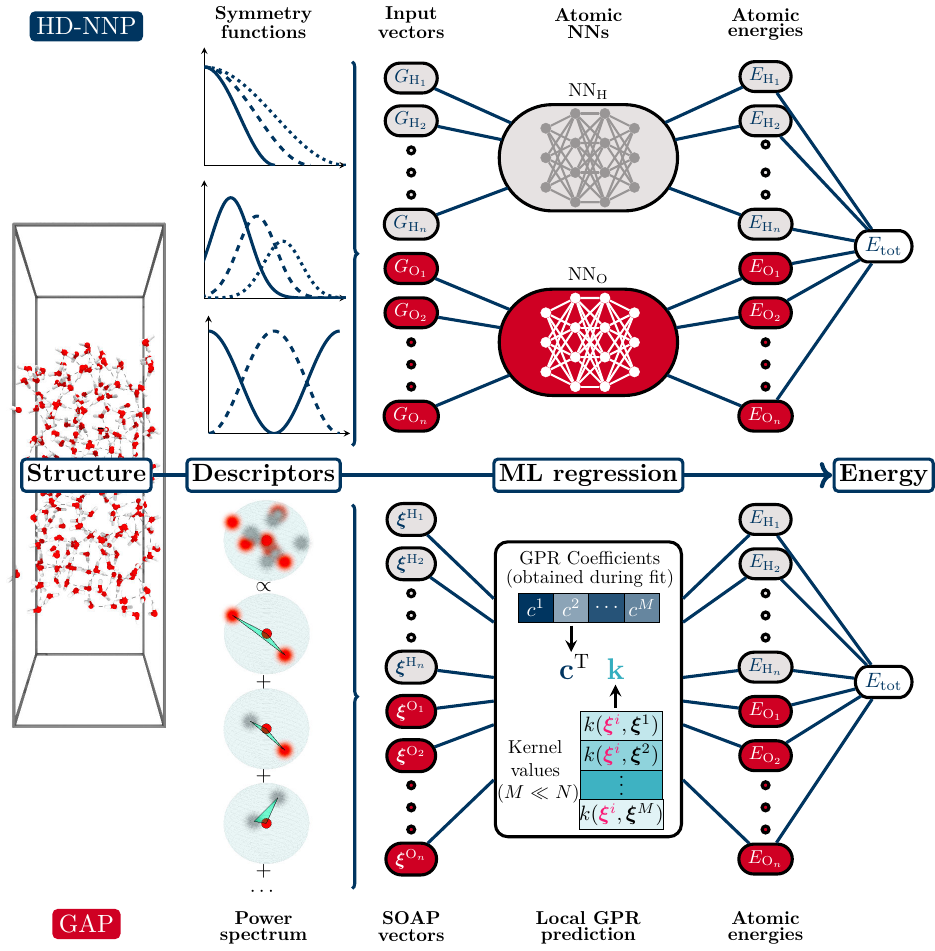}
\caption{%
\textbf{Representation of the structure--energy relation realised
by high--dimensional neural network potentials (HD-NNP, top)
and Gaussian approximation potential (GAP, bottom) for the
description of a slab of water.}
In the first step, the structure is transformed either 
via atom-centred symmetry functions (HD-NNP) or
smooth overlap of atomic positions (GAP) into rotationally,
translationally, and permutationally invariant vectors.
These serve as the input for atomic neural networks (HD-NNP) or
local Gaussian process regression models (GAP) to provide atomic energies
that sum up to the total energy of the system. 
This functional relation is analytically differentiable and, thus, can also 
provide the interatomic forces.
\label{fig:hdnnp-gap}
}
\end{figure*}

\subsubsection{Gaussian Approximation Potentials}
\label{sec:gap}

Having covered the general idea behind GPR, let us now apply these concepts
to MLPs by discussing the Gaussian Approximation Potential
(GAP) methodology developed by Bartok \textit{et al.} \cite{Bartok2010}.
As in many conventional and machine learning-based potentials, the key assumption
behind GAP is that the potential energy $E_{\mathrm{tot}}$ is constructed as a 
sum of the atomic energies, $\epsilon_i$, such that
\begin{equation}
    E_{\mathrm{tot}} = \sum_i^{N_\mathrm{atom}} E_i \mbox{ , }
    \label{eq:GAP_sum_potE}
\end{equation}
where $N_{\mathrm{atom}}$ corresponds to the number of atoms in the system.
An illustration of the resulting structure-energy relation is shown in the lower part of Fig.~\ref{fig:hdnnp-gap}.
To account for the distinct energy and length scales of the repulsive and 
the attractive regime of potential energy surfaces, atomic energies, 
$E_i$, are computed based on a linear combination of $d$-dimensional terms.
Most commonly, this involves a two-body (2B) and many-body (MB) term which
are both expressed as separate GPR models using suitable descriptors,
$\textbf{q}^{(d)}$, such that
\begin{equation}
    E_{\mathrm{tot}} = \delta^{(2B)} \sum_{ij} E^{(2B)} (\textbf{q}_{ij}^{(2B)}) +  \delta^{(MB)} \sum_{i} E^{(MB)} (\textbf{q}_{i}^{(MB)})  \mbox{ , }
    \label{eq:GAP_sum_potE_terms}
\end{equation}
where each term is weighted by the scaling factors $\delta^{(d)}$
which represents an additional hyperparameter.
In many GAP models \cite{Deringer2017, Rowe2018,Rowe2020/10.1063/5.0005084,Rowe2022/10.1063/5.0091698/2841205, Thiemann2020},
the two-body term makes the largest contribution to the total energy, 
thus the $\delta^{(2B)}$ is usually set to be five to fifty times larger
than $\delta^{(MB)}$.
The squared exponential or Gaussian kernel is commonly employed as the
kernel function for the low-dimensional 2B term, using the distance
between two atoms as a descriptor, resulting in a sparse GPR model 
using $M^{(2B)}$ representative points.
The MB term, conversely, is represented by the SOAP kernel, defined in 
equation \ref{eq:SOAP_kernel}, using a set of $M^{(MB)}$ representative
configurations resulting in the following expression for the potential energy
\begin{align}
    E_{\mathrm{tot}} = &\delta^{(2B)} \sum_{ij} \sum_{m=1}^{M^{(2B)}} c^{(2B)}_m \exp \left[ - \frac{|R_{ij} - R_m |^2}{2\theta^2} \right] \nonumber\\
    &+  \delta^{(MB)} \sum_{i} \sum_{m=1}^{M^{(MB)}} c^{(MB)}_m  (\bm{\xi}_i\cdot \bm{\xi}_m)^\zeta \mbox{ , }
    \label{eq:GAP_sum_potE_terms_explicit}
\end{align}
where $\theta$ is a hyperparameter related to the 2B descriptor, 
and $\bm{\xi}$ is the normalised power spectrum vector as outlined 
in the SOAP section above.
The coefficients of the individual GPR models, $c^{(d)}_m$, are obtained 
during the fitting process outlined in the previous section.
In order to select the representative points or basis functions for a descriptor,
there are different approaches depending on the dimensionality of the descriptor. 
For low-dimensional descriptors, such as the 2B descriptor, a uniform grid 
in the one-dimensional space is chosen to ensure all interatomic distances
are well represented, resulting in a relatively small number of representative
points, typically less than 100.
On the other hand, high-dimensional descriptors, such as the SOAP representation,
require a more efficient strategy, such as using the so-called CUR algorithm, which 
maximises the span of the basis set in a low-dimensional representative subspace of
the full descriptor set. 
The exact number of representative points needed depends on the complexity
of the target phase space region the model is being fitted to, and is usually around
several thousand.

The regularisers, or data-point-specific variances, $\sigma_n$, are an important
set of parameters in determining the coefficients, $c_m^{(d)}$,
for the GPR model.
These regularisers need to be chosen carefully for each data point in
the training set, as too large of a value will result in poor agreement 
with the reference data, while too small of a value can lead to overfitting.
It is important to keep in mind that the reference data, such as energies,
forces, and stresses, may contain noise due to unconverged electronic structure
calculations or other issues related to the \textit{ab initio} reference method.

The inherent approximation of decomposing the potential
energy into local contributions also represents a potential bias.
The appropriate choice of the regulariser, $\sigma_n$, depends on multiple
factors such as the property being modelled (energy, force, or virial stress)
and the location in the phase space of the configuration. 
Being closer to the potential energy minimum, solid configurations 
require more precise fits than liquid configurations,
and typical values are ($\sigma_E = 0.001$, $\sigma_F = 0.05$ , $\sigma_V = 0.05$)
and ($\sigma_E = 0.03$, $\sigma_F = 0.2$ , $\sigma_V = 0.2$) for a solid and
liquid configuration, respectively, with units of eV/atom for energies
($\sigma_E$) and virial stresses ($\sigma_V$) and eV/Å for force 
components ($\sigma_F$).
\subsection{Discussion and Outlook of Regression Models}
\label{sec:reg-out}
We have now described two routes -- HD-NNPs based on artificial neural networks and GAPs based on Gaussian process regression -- that relate structural descriptors to forces and energies.
These can now be used to extend both the length and time scales of atomistic simulations, with the caveat of any new chemical environments remaining close to those in the training data.

The differences between these two methods can be broadly classified with respect to their descriptors and model complexities.
The ACSF descriptors for HD-NNPs are typically low-dimensional descriptors which are then transformed via a highly flexible NN.
In contrast, the SOAP descriptors required for the GAP models are of much higher dimension, but this makes the actual regression task ultimately simpler.

These models can be further compared in terms of similarities and differences, as well as their advantages and disadvantages.
A task common to the training of any machine learning model is the judicious selection of the hyperparameters for the model.
These need to be chosen before the model training.
A combination of experience and systematic testing is the typical approach.
An immediate advantage of the kernel models is that they have an inherent uncertainty estimate available, while NN approaches require additional effort to obtain this.
For example, multiple NN-based models can be trained and combined into so-called committee models, where the ensemble variation can provide a powerful uncertainty estimate.~\cite{Schran2020/10.1063/5.0016004}
In terms of computational efficiency, GAP can be expensive for large datasets compared to HD-NNPs.
The cost of the regression task scales as \rev{$\mathcal{O}(NM^2)$} to \rev{compute the coefficients}.
Evaluation of both HD-NNP and GAP models scale linearly with the number of model parameters but for GAP the number of model parameters is related to the number of representative points in the training data and so overall scales $\mathcal{O}(M)$.
This makes NNPs a more attractive option for large and complex (heterogeneous) data sets which require a larger number of representative points.
Alternatively, recent developments in another family of methods expand the PES in terms of many-body correlation functions.
While both NNPs and GAP rely on low-body order descriptors, moment tensor potentials \cite{Shapeev2016/10.1137/15M1054183} and the atomic cluster expansion (ACE)\cite{Drautz2019/10.1103/PHYSREVB.99.014104}, are computationally efficient up to high-body order and allow treatment of complex chemical environments involving many different elements.
Another general disadvantage of both HD-NNPs and GAP models is the inherent assumption of locality, where the models are truncated at a finite interaction distance.
This can lead to significant errors for systems in which intrinsically long-range electrostatic or dispersion interactions are important \cite{Ko2021/10.1021/ACS.ACCOUNTS.0C00689}.
\rev{In Section~\ref{sec:devel}, we will discuss methods for overcoming these challenges in more detail.}

In the end, the choice of the flexibility in the descriptors and regression model is 
a smooth scale, ranging from the most extreme cases of having all flexibility in the descriptors
as in the ACE \cite{Drautz2019/10.1103/PHYSREVB.99.014104} that use linear regression, to message-passing NNs \cite{Zubatyuk2021/10.1038/s41467-021-24904-0, Zubatyuk2019/10.1126/sciadv.aav6490} that use rather simple descriptors \rev{and are introduced in more detail in the next section}.
In practice, the optimal solution has not yet been found and computational efficiency
also needs to be taken into account.
A useful way of understanding differences in this respect is by looking at the so-called
Pareto front of accuracy-versus-cost achieved with different models.
\rev{
Cost is then usually measured as the time per simulation step using the model,
not taking into account the development time and training cost.
}
The best models will provide the lowest cost with the highest accuracy over a large front
compared to other available models and architectures.
We refer to further aspects on the assessment of the accuracy of a given model in section~\ref{sec:validation}.
A notable development towards higher efficiency is the use of graphics processing units
acceleration.~\cite{Bochkarev2022/10.1103/PhysRevMaterials.6.013804,
Fan2022/10.1063/5.0106617}
This is expected to become more and more widespread further
pushing the boundaries of what can be done with MLPs.
\rev{It is hard to provide a comprehensive recommendation as to which model is best for specific systems.
However, the push towards open source and automated development as well as standardised formats and validation tests, as summarised in Tab.~\ref{tab:code}, enables users to easily compare different architectures.
We hope that the overview of current developments as discussed in the next section will provide a good starting point for choosing a suitable model for new work.
}

\section{Current Developments}
\label{sec:devel}
After understanding how the seminal works of high-dimensional
MLPs solved the challenges associated with representing PESs, we
can now widen the view on the field and look into more recent
developments.
Overall, these developments either improve upon the chemical
descriptors, or the architecture of the regression, targeting improved accuracy, efficiency and generalisation.
Another important development is finding an optimum in
terms of speed and accuracy.
So far the models we have discussed comprise a set of predefined truncated descriptors which are then used as input for a highly non-linear function, the output of which then implicitly contains higher body-order correlations. 
However many modern MLP approaches instead aim to explicitly capture these correlations.

As previously discussed in earlier sections, integral to all ML models is the consideration of the physical symmetries of the system.
In general, all of the properties of a particular atomic structure obey a general symmetry constraint:
\begin{equation}
    \phi\{Q\cdot\sigma_i\} = Q\phi\{\sigma_i\}
\end{equation}
This shows that operating on an atomic configuration ($\sigma_i = r_{1i}, r_{2i},..., r_{Ni}$) with a symmetry operator $Q$ (e.g. translation, rotation) and then taking the ML model output, $\phi\{\ldots\}$, should be equivalent to operating on the ML output of the original atomic configuration $\sigma_i$ and applying the symmetry operation after.
In general, different physical properties can transform differently under specific symmetry operations.
In atomistic modelling, the focus lies on the Euclidean symmetries (translations, rotations, and reflections) of three-dimensional space, represented by the E(3) group.
Since translational invariance is maintained by working with interatomic distances rather than positions, our attention is primarily on rotations and reflections, forming the O(3) group.
Upon applying a symmetry operation from this group, a property can either remain invariant or transform equivariantly.
For instance, scalar properties such as the global potential energy, which will not change for example if a molecule is rotated (in the absence of an external field), are invariant with respect to O(3) operations.
Vectors or higher-order tensorial properties such as forces and dipole moments, conversely, should obey the same rotation. 
Also, as previously mentioned in Section~\ref{sec:desc_outlook}, the descriptors should be capable of unambiguously differentiating atomic environments, and so should be formally complete \cite{Pozdnyakov2020/10.1103/PHYSREVLETT.125.166001}.
Recent work can therefore be roughly categorised into four main objectives, 1) generation of a complete set of descriptors, 2) incorporating the representation of the atomic environments directly in the model architecture, which are then another learnable feature, 3) going beyond a local description of atomic environments, and 4) providing generalisable models across large regions of compound space.%

\subsection{Completeness of Descriptors}
\label{subsec:ace}

In general, the energy of an atom $i$ can be systematically written as a many-body expansion:
\begin{equation}\label{eq:ACE_expansion}
    E_i = V_1(r_i) + \frac{1}{2} \sum_{j} V_2(r_{ij}) + \frac{1}{3!} \sum_{j,k}V_3(r_{ijk}) + ...
\end{equation}
where each term successively depends on an additional particle.
However, this expansion scales very poorly with increasing leading body order $v$ ($\mathcal{O}N^v$) for $N$ neighbours of atom $i$ within a cutoff and is computationally intractable to go beyond body order $v=5$.
Moment tensor potentials \cite{Shapeev2016/10.1137/15M1054183} and more recently the atomic cluster expansion (ACE) \cite{Drautz2019/10.1103/PHYSREVB.99.014104} overcome this issue to give efficient linear-scaling models \cite{Novikov2020/10.1088/2632-2153/ABC9FE, Lysogorskiy2021/10.1038/s41524-021-00559-9}.
Similar to using single particle orbitals in quantum chemistry and building a Slater determinant, ACE generates a basis $A_{iv} = \sum_j \phi_v (r_{ij}))$ of one-particle functions ($\phi_v()$) to describe the local atomic environments.
This basis is permutationally invariant as a result of summing over all neighbours and is complete.
The energy in equation \ref{eq:ACE_expansion} is then constructed from products of these 1-particle basis functions $A_{iv}$:
\begin{align}
    E_i &= \sum_v c^{(1)}_v A_{iv} + \sum_{v1v2}^{v1\geq v2} c^{(2)}_{v1v2} A_{v1}A_{v2} + \\ &\sum_{v1,v2,v3}^{v1\geq v2 \geq v3} c^{(3)}_{v1v2v3} A_{v1} A_{v2} A_{v3} + ...
\end{align}
By averaging the $A$ basis over rotations (see Ref.~\citenum{Drautz2019/10.1103/PHYSREVB.99.014104}), the total energy can be written as a polynomial of a new rotationally invariant $B$ basis functions:
\begin{equation}
    E = \sum_{vi} c_{iv}B_{vi}
\end{equation}
This resulting polynomial is linear-scaling ($\mathcal{N}$) irrespective of the body-order of the expansion \cite{Kovacs2021/10.1021/acs.jctc.1c00647}.
ACE is a complete expansion of the atomic environment and can be used as a framework to classify other types of potentials. 
For example, moment tensor potentials introduced already in 2016\cite{Shapeev2016/10.1137/15M1054183} are also a complete basis spanning the space of atomic environments and have a 1:1 mapping to ACE \cite{Batatia2022Design}.

\subsection{Learnable Descriptors: Graph Neural Networks}
\label{subsec:mess}
So far, all of the MLP models have been based on (systematic and complete in the case of ACE and MTPs) fixed descriptors of the local atomic environment.
However, another class of MLPs are based on the well-established field of graph neural networks (GNNs), where the representation can be learned from molecular graphs.
GNNs exploit the fact that, in general, atomic environments are highly amenable for representation as graphs, where nodes (atoms) are connected to all other nodes (within a cutoff) via edges $e_{ij}$.
For the successful subclass of message-passing neural networks (MPNNs),
each atom/node $i$ is associated with a latent state $h_i$, updated with each message-passing iteration $t$.
A message $m_i^{t+1}$ is constructed on a node $i$ by `pulling' information of the states from all neighbouring atoms $N(i)$ using a message function $M_t$:
\begin{equation}
    m_i^{t+1} = \sum_{j \in N(i)} M_t(h_i^{(t)}, h_j^{(t), e_{ij}})
\end{equation}
The state of each node is then updated based on these messages
\begin{equation}
    h_i^{t+1} = U_t(h_i^t, m_i^{(t+1)})
\end{equation}
where $U_t$ is a learnable node update function.

Early architectures such as Schnet\cite{Schutt2018/10.1063/1.5019779} and Dimenet\cite{Gasteiger2020/2003.03123} were based on invariant features $h_i$.
This guarantees that the predicted energy will be invariant under the symmetry operations described.
More recent MPNNs are instead based on vectorial or tensorial representations at each node.
These models include NequIP \cite{Batzner2021/10.1038/s41467-022-29939-5}, PaINN \cite{Schutt2021/2102.03150},  NewtonNet \cite{Mojtaba2022/10.1039/D2DD00008C}, and SEGNN \cite{Brandstetter2022}.
Coupling these ideas with attention and transformer-like concepts leads to another class of models~\cite{Fuchs2020,Liao2023,Tholke2022/2202.02541,Passaro2023}.
All these architectures are inherently \textit{equivariant} with respect to the relevant symmetries of rotation, translation and inversion (E3 symmetry group).
Such equivariant MPNNs are much more data efficient since the relevant symmetries are already encoded in the model, and therefore large amounts of data are not required to `learn' these equivariances and have greater accuracy.
Unlike the NNP and GAP models discussed in previous sections, which strictly contain local chemical information, MPNNs can propagate semi-local information via iterative message-passing steps.
Thus, the so-called `receptive field' $r_{c,e}$ of an MPNN is expanded beyond the local atomic cutoff $r_{c,l}$ based on the number of message passing layers $N_l$: $r_{c,e} = N_t r_{c,l}$.
Information about increasingly non-local features can be built up through multiple message-passing layers $t$. 
However, this expanded receptive field of equivariant MPNNs leads to a significant scaling problem, making parallelisation involving multiple message-passing steps very difficult since there is a cubic scaling of the number of neighbouring atoms with the number of message-passing steps.
For example Ref.~\citenum{Musaelian2023/10.1038/s41467-023-36329-y}
shows that an MPNN for bulk water with a local cutoff of 6\AA containing 96 neighbours increases to 20834 upon 6 message passing steps.
This highlights that maintaining the locality of the model is highly desirable to allow for efficient implementation of these methods.

The relationship between overall body order of the features $h_i$ and message passing has been explored in several recent works \cite{Batatia2022/2205.06643, Bochkarev2022/10.1103/PHYSREVRESEARCH.4.L042019, Nigam2022/10.1063/5.0087042/2841327}.
While the previous equivariant models discussed exploited message passing, they only considered 2-body messages, resulting in lower efficiency due to the increased message-passing steps required to capture many-body correlations.
The most recent state-of-the-art methods -- Allegro \cite{Musaelian2023/10.1038/s41467-023-36329-y} and MACE \cite{Batatia2022/10.48550/arxiv.2206.07697} -- thus combine message passing with high body order features.
For MACE, each layer now comprises many-body messages, resulting in efficient potentials that can be systematically expanded to arbitrary body-order, thereby explicitly including higher-order correlations, without requiring the many message passing steps of previous MPNNs (typically 2 compared to ~6).\cite{Batatia2022/10.48550/arxiv.2206.07697}
MACE has recently been shown to provide convincing accuracy across application in many diverse areas~\cite{Kovacs2023/10.1063/5.0155322}.
Other MPNNs have be very useful for universal potentials across the periodic table~\cite{Chen2022/10.1038/s43588-022-00349-3}.
In summary, MPNNs are accurate, data-efficient, and fast and are expected to be at the forefront of next generation MLP applications and developments.

\subsection{Beyond Locality} %
\label{subsec:phys}
While elemental systems and those with significant screening effects \cite{Yue2021/10.1063/5.0031215} can be successfully treated with short-range models, such models can fail for systems governed significantly by long-range electrostatic or dispersion interactions.
Therefore the inherent locality assumption of MLPs --required to facilitate practical implementation -- is a major shortcoming.
Some models include physically motivated corrections -- similar to DFT -- for example, PhysNet \cite{Unke2019/10.1021/ACS.JCTC.9B00181} and Tensormol \cite{Yao2018/10.1039/C7SC04934J} use Grimme's DXX family of dispersion corrections \cite{grimme2006/10.1002/JCC.20495}. 
However, many approaches to address this issue are based on a decomposition of the total energy as a sum of short-range terms $\mathrm{E_{sr}}$ and long-range $\mathrm{E_{elec}}$ contributions:
\begin{equation}\label{eq:3gnnp}
    E_{tot} = E_{sr} + E_{elec}
\end{equation}
The most straightforward approaches simply involve subtraction of the long-range electrostatic component using a standard Ewald-like sum \cite{Bartok2010, Deng2019/10.1038/s41524-019-0212-1} and subsequently training a purely short-range model to the difference.
However, this assumes fixed charges and also raises the question of what is the correct partial charge to assign to the atoms.
This can be addressed by exploiting machine learning to address the electrostatic component, thereby training two models for the short and long-range components.
So-called 3rd generation MLPs account for environment-specific charges, by training an additional NNP to predict the atomic charges or higher order multipoles as a function of the chemical environment \cite{Artrith2011/10.1103/PhysRevB.83.153101, Bleiziffer2018/10.1021/ACS.JCIM.7B00663, Ramakrishnan2015/10.1021/acs.jctc.5b00099, Nebgen2018/10.1021/ACS.JCTC.8B00524, Yao2018/10.1039/C7SC04934J, Metcalf2021/10.1021/ACS.JCIM.0C01071, Zubatyuk2021/10.1038/s41467-021-24904-0, Unke2019/10.1021/ACS.JCTC.9B00181}.
These models aim to reproduce reference partial charges on atoms in a system, which are then used in an Ewald summation to compute the long-range energy contribution.
Since charges are not a quantum mechanical observable, some charge partitioning schemes should be used on the electronic structure calculations to obtain the reference charges.
Examples include Hirschfeld and Bader charges or from dipole moments -- however, it should be noted that the choice of charge partitioning scheme is not unique.
These are then subtracted from the reference energies and forces so to avoid double-counting portions of the long-range term, and a standard HDNNP is trained to give the short-range contribution $E_{sr}$.
While this third-generation NNP approach has been successful in capturing long-range and even dispersion interactions, there is still an issue in describing non-local effects.
There are many cases in chemistry and biology in which there is a global change in the electronic structure such as long-range charge transfer or when a system has multiple charge states, which occur outside of the local chemical environment considered by the cutoff or message passing steps.
For example, protonation/ deprotonation of a molecule overall results in a change in the total charge of the system.
Other situations for example in surface science where dopants far from the adsorbate influence adsorption geometry and binding again require a model that can faithfully capture interactions which extend far beyond the typical cutoff.
So-called 4th-generation models thus use atomic charges based on the global charge of the system.
A charge equilibration scheme \cite{Ko2021/10.1038/s41467-020-20427-2} redistributes the charge density over the system to minimise the total electrostatic energy. 
Similar to equation \ref{eq:3gnnp}, the total energy is again split into short and long-range contributions, however now the short-range part contains non-local charge information obtained via charge equilibration:
\begin{equation}
    E_{total}(R,Q) = E_{elec}(R,Q) + E_{short} (R,Q)
\end{equation}
Other promising models are based on a more global approach such as the long-distance equivariant representation (LODE) \cite{Grisafi2019/10.1063/1.5128375}, where non-local information is explicitly incorporated into the feature descriptors and symmetric gradient-domain machine learning (sGDML) \cite{chmiela2023/10.1126/SCIADV.ADF0873} which directly learns the forces to give an energy-conserving global model.
\subsection{General Purpose and Foundational Models}
\label{subsec:found}
Another very promising recent development is the establishment of MLPs that are applicable throughout chemical compound space by training on very large and diverse datasets.
Initially, such attempts were mostly limited to property predictions and minimum energy structures since achieving robust and accurate force and energy predictions for stable simulations had been prohibitively difficult.
For deeper insights into the concepts of property predictions across chemical compound space, we refer the interested reader to the expert review by Huang and von Lilienfeld~\citenum{Huang2021/10.1021/ACS.CHEMREV.0C01303}.
Earlier attempts at generalisable representations of PESs have concentrated on relatively well-defined regions of compound space, such as elementary systems like carbon~\cite{Rowe2020/10.1063/5.0005084}, or silicon~\cite{Deringer2021/10.1038/s41586-020-03072-z} showcased in section~\ref{sec:examples} in more detail.
The recent progress in model architectures, data curation, and training algorithms has led to recent successes in delivering MLPs that provide stable simulations across very diverse systems and beyond their training domain.
Examples of these advanced models include MACE-MP-0~\cite{batatia2023foundation}, MACE-OFF~\cite{kovacs2023maceoff23}, GNoME~\cite{Merchant2023/10.1038/s41586-023-06735-9}, MatterSim~\cite{yang2024mattersimdeeplearningatomistic}, and CHGNet\cite{deng2023chgnet}.
These models make a promise of being foundational models for molecular and materials modelling.
They have demonstrated the capability to generalise well beyond the systems they were initially trained on, enabling accurate and reliable simulations for a wide range of chemical compounds and materials.
One of the significant advancements contributing to this success is the development of more sophisticated neural network architectures that can capture the intricate details of inter-atomic interactions.
Additionally, the creation of extensive and high-quality training datasets, such as the MPtrj dataset\cite{deng2023chgnet}, which encompass a broad spectrum of chemical environments, has been crucial.
This comprehensive data helps the models to learn more generalised features, leading to improved performance across various types of compounds.
Moreover, the improvement in training algorithms, including techniques to prevent overfitting and methods to ensure the physical plausibility of the predictions, has further enhanced the robustness of these models.
Another very promising direction of this research is the possibility for fine-tuning the model to new training data and other reference methods by utilising the concepts of transfer learning.
Techniques like transfer learning, or delta learning can be employed effectively, where a pre-trained model is fine-tuned with new training data or adapted to different reference methods, thus extending its applicability and improving its accuracy for specific tasks.
This approach allows the foundational models to be adapted for specialised applications, making them incredibly versatile and powerful tools for the molecular and materials modelling community.
By continuously integrating new data and refining their algorithms, these models are expected to evolve, providing even greater accuracy and efficiency in simulations, ultimately accelerating the discovery and development of new materials and chemicals.

\section{Data Set Generation}
\label{sec:data_gen}
After having seen how a robust and accurate representation of the PESs
can be achieved with both kernel and NN-based methods, we will now
concentrate on the generation of representative data sets.
Each machine-learning model is only as good as its underlying data.
Thus, special care should be put into curating representative and well-balanced data.
In addition, reference calculations can become expensive --- in particular
when thinking of correlated wavefunction methods --- so additional
emphasis should be put on keeping data sets compact.
When thinking about this task, one might simply start out with
generating random configurations and computing their energies (and forces) with a suitable
reference method, such as DFT.
However, in most cases, this will result in very unfavourable structures,
so some physical sampling in preparation for the data set generation
is usually advantageous.
This can either be done by molecular dynamics or any other sampling
technique applicable to the specific system of interest.
Generally, the optimal approach will depend on the application the user has in mind and can vary significantly from system to system.
For instance, including data corresponding to local minima of the PES will be crucial for the model to reliably predict the equilibrium configuration of a solid while data at higher temperatures will be required to describe a liquid phase.
Next, one might start selecting a random sample from the previously
sampled pool of configurations, thus generating a Boltzmann-weighted
distribution.
\rev{
These random sampling techniques, and also other hand-crafted selection methods,}
can in many cases result in a decent first model, but remain
relatively ad hoc.
\rev{Overall,} it is usually quite easy to generate new configurations, but it is a much harder task to select the ``right'' ones.

Fortunately, there are multiple strategies to generate data sets in a
more data-driven and automated way which will be presented in the
following sections.
These can be roughly categorised based on which side of the structure-energy relation they are operating on.
There are multiple strategies that rely on structural similarity measures
to select new configurations to be added to the data set.
Furthermore, so-called uncertainty based techniques can be used to
select points where the model shows a high uncertainty in its prediction.
\subsection{Structural Selection Techniques}
\label{sec:fps}

A popular similarity-based selection technique is
farthest point sampling (FPS) \cite{De2016,Bartok2017/10.1126/SCIADV.1701816}, which is a greedy algorithm
that selects the structure that is most different from the previously chosen
structures, in order to create a structurally diverse training set.
To measure the similarity between two structures, $A$ and $A'$, FPS employs
local representations $\textbf{q}_{i}^{(d)}$ to describe the atomic 
environments in d-dimensional descriptor space, such as those discussed above.
The most intuitive approach to use these local descriptors to compare
and match entire configurations by averaging the individual representations
$\textbf{q}_{i}^{(d)}$ over all atoms in each configuration, resulting in
a mean representation of each structure, $\bar{\textbf{q}}^{(d)}$.
By comparing $\bar{\textbf{q}}^{(d)}_{A}$ and $\bar{\textbf{q}}^{(d)}_{A'}$,
the topological difference between the configurations $A$ and $A'$ can be quantified. 
This similarity is represented by a distance, $L$, and FPS selects 
configurations such that the distance between a new configuration,
$\bar{\textbf{q}}_{m+1}^{(d)}$, and all previously chosen structures in 
the database, $Q_\mathrm{chosen} = \{\bar{\textbf{q}}_{1}^{(d)},
\bar{\textbf{q}}_{2}^{(d)}, ..., \bar{\textbf{q}}_{m}^{(d)} \}$,
is maximized, as shown in the equation:
\begin{equation}
\bar{\textbf{q}}_{m+1}^{(d)} = \mathrm{argmax}_{ \bar{\textbf{q}}^{(d)}} [L(Q_\mathrm{chosen}, \bar{\textbf{q}}^{(d)})] \mbox{ , }
\label{eq:FPS}
\end{equation}
where $\bar{\textbf{q}}^{(d)}$ is the structurally averaged descriptor
for all configurations in the pool of potential training structures.
It is possible to use different distance metrics $L$ such as the summed
Euclidean distance of the query points to all points in the existing data set.
The SOAP descriptor is particularly useful in this context as
the dot product of two independent SOAP descriptors corresponds directly to the 
overlap and structural similarity.~\cite{Jager2018/10.1038/s41524-018-0096-5}
However, other descriptors such as ACSFs can also be used as long as sufficient
distance measures are introduced.

Another technique that operates directly on structures is normal mode sampling~\cite{Smith2017/10.1039/C6SC05720A}.
Here, new structures are generated by using the normal modes of an equilibrium geometry to generate distorted structures along the normal modes according to the thermal harmonic oscillator distribution.
This technique has recently been extended to consider structures along
reaction pathways in so-called transition tube sampling.~\cite{Brezina2023}

\subsection{Active Learning}
\label{sec:active}
\begin{figure*}[tb]
    \centering{}
    \includegraphics[width=1.0\textwidth]{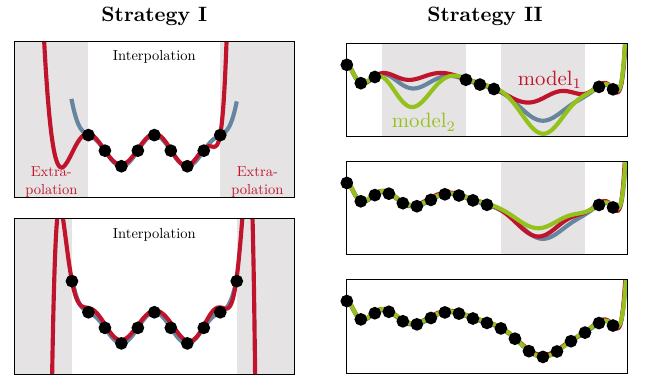}
    \caption{
        \textbf{Representation of two general strategies for an
        improvement of a machine learning potential \rev{applicable to both neural network and Kernel-based approaches}.}
        Left: Improvement of the boundaries of the reference
        set can be achieved by adding structures that where
        detected to leave the range of descriptors
        encountered in the training set.
        Right: Regions that are underrepresented in
        the training set can be improved by adding
        structures to the training set for which two
        slightly different models provide diverging
        predictions.
        The reference potential energy surface is shown in blue,
        the machine learning models in red and green,
        and the training points of the reference function
        are depicted as black dots.
        Regions that are not well represented by the model are
        highlighted in grey.
    }
    \label{fig:nnp_imp}
\end{figure*}

The main advantage of machine learning approaches over traditional functional forms
is their flexibility and the possibility for iterative improvement.
While low dimensional potentials in general do not have to become
better after a certain point when more reference data is taken into
account, machine learning models will gradually improve, if additional
data is included in the fitting procedure.
Due to the flexibility, however, structures far away from any point
in the training set are not well represented
and are therefore prone to extrapolation errors.~\cite{Gubaev2018/10.1063/1.5005095}
The set of structures used for the training therefore has to reflect 
the relevant configurations encountered in the subsequent application.
In general, two different scenarios for extrapolation problems are
possible.
The first problematic situation is a region that is beyond the boundary
of the configuration space spanned by the training set as sketched
on the left side of Fig.~\ref{fig:nnp_imp}.
If in such a situation the structure is very far away from the boundaries,
the model loses all its predictive power, due to the absence
of physical insight.
These situations are fortunately easy to detect by 
comparing the descriptors of a configuration
in question to the range of values encountered in the training set.
If any descriptor is outside of this range the new
structure suffers from extrapolation and will most likely not be
represented with sufficient quality.
The second case in which MLPs will not provide
plausible predictions is for regions that are 
underrepresented in the training set as shown on the right side
of Fig.~\ref{fig:nnp_imp}.
These cases are much harder to detect for a single model,
but can be prevented by appropriate preparation of the training set
to avoid holes.
In addition, two models fitted with different starting
parameters will provide very different predictions in such a region,
due to the large quantity of local minima in parameter space.
By comparison of the prediction of two slightly different models
it is therefore possible to detect exactly such regions without
ever inquiring the usually expensive reference method.
\rev{Gaussian Process based models can easily identify both situations using their intrinsic error estimate that will flag regions far away from any other data.}

At first, the poor capability of machine learning models for extrapolation
might look like a significant drawback of these methods, since
any transferability to unknown situations is lost and
predictions become useless.
But at the same time, these properties allow for
very powerful strategies to improve the description
of the model and automate the process of assembling
the reference set.
As shown in Fig.~\ref{fig:nnp_imp}, when points are
iteratively added to the training set that
have been detected to be either outside the boundaries
or in underrepresented regions of the configuration space
spanned in the training set, these regions can be
selectively improved.
If combined with a physically motivated sampling of the
underlying PES, 
this can be used to generate structure-energy relations
in an unbiased and highly efficient way.

These concepts can be generalised under the framework of
active learning, where the most suited configurations
for an improvement of the model are added to the training set.~\cite{Gastegger2017/10.1039/C7SC02267K,
Podryabinkin2017/10.1016/J.COMMATSCI.2017.08.031,
Smith2018/10.1063/1.5023802,
Deringer2018/10.1103/PHYSREVLETT.120.156001,
Zhang2019/10.1103/PHYSREVMATERIALS.3.023804,
Musil2019/10.1021/acs.jctc.8b00959,
Zhai2020/10.1063/5.0002162,
Jinnouchi2020/10.1063/5.0009491,
Lin2020/10.1063/5.0004944}
The term originates from the idea that the learning algorithm 
can interactively query an ``oracle'' to label new data points
with the desired outputs.
This is usually done by having access to some kind of
uncertainty estimate, allowing to filter large sets of potential
candidate structures which do not have to be labeled with the
respective reference method.
One can then design iterative procedures which cycle through
the steps of 1) getting an uncertainty estimate for a large pool
of candidate structures, 2) selecting and labelling a small set of
structures with the highest uncertainty, and 3) training an improved model.

In the context of neural network based MLPs, a single model usually
not provide an uncertainty estimate (although there are architectures where
this can be achieved\cite{Janet2019/10.1039/C9SC02298H,
Carrete2023/10.1063/5.0146905/2891492}).
However, we have seen that comparing the prediction of two slightly different
models can provide us with an indication of uncertainty.
This can be formalised in so-called ensemble or committee models, where
multiple HD-NNPs are combined and the committee members are separately
trained from independent random initial conditions to a subset of
the total training set.~\cite{Schran2020/10.1063/5.0016004}
While the committee average provides more accurate predictions than
the individual HD-NNPs, the committee disagreement, defined as the
standard deviation between the committee members, grants access
to an estimate of the error of the model.
This committee disagreement provides an objective measure of the
\rev{error} of the underlying model\cite{Imbalzano2021/10.1063/5.0036522}.
To construct a training set of such a model in an automated
and data-driven way, new configurations with the highest disagreement
can be added to the training set.
This is an active learning strategy called query by committee (QbC)
and can be used to systematically improve a machine-learning model.~\cite{Krogh1994, Seung1992/10.1145/130385.130417}
This has been utilised extensively in recent times for the automated
development of NNPs for various systems.
As seen above, MLPs based on GPR have a built-in uncertainty estimate and can be
used for similar active learning strategies.
\subsection{Reinforcement Workflows}
\label{sec:workflows}

The above-described data-driven techniques for the selection of
new points to be added to a training set enable the user to
establish workflows for the improvement of an MLP.
It has become standard to train an initial model, which is
subsequently reinforced to better reproduce user-selected and
problem-specific properties, or expand into regions of phase
space that were previously not part of the training data.
Initial models can often be used as very effective structure
generators, thus preventing the requirement for expensive
reference calculations during sampling.
Either structure-based, or uncertainty-based criteria, or
a combination of both are then used to filter the large set
of structures and label the data.
Recent developments have shown that it is possible to use
the uncertainty estimate of MLPs either to stabilise 
a simulation in regions further away from the existing data~\cite{Schran2020/10.1063/5.0016004},
or bias simulations towards higher uncertainty for quicker
exploration of configuration space~\cite{Oord2022,
Kulichenko2023/10.1038/s43588-023-00406-5}.
\rev{
The latter has been introduced under the term
``hyperactive learning'' as it enables faster generation
of configurations for an improvement of the model.
}
\section{Validation}
\label{sec:validation}

After having seen how structure-energy relations can
be represented with MLPs and how data sets can be assembled,
we need to focus on the validation of the resulting model.
This is one of the most important steps in the development
of a new model in order to verify that we can trust the
predictions of the MLP.
For this tutorial, we will mostly focus on two types of validation steps,
numerical errors with respect to the ``learned'' properties,
and secondary properties derived from the representation of the PES.
For further details, we refer the interested reader to Ref.~\citenum{Morrow2022/10.1063/5.0139611}
which provides an excellent detailed introduction to this topic.

\subsection{Primary Properties and Numerical Errors}
\label{sec:prim_prop}
A starting point to assess the performance of a machine learning model
is to use metrics like Root Mean Squared Error (RMSE)
and Mean Absolute Deviation (MAD) for the primary properties of the model.
These are the reference energies $E$ and forces $F$ that the model is trying to
reproduce directly.
RMSE is a commonly used metric that measures the average difference
between predicted and actual values.
It is calculated as the square root of the average of the squared
differences between the predictions and the actual values, e.g. for energies $E$
\begin{align}
    E^\text{RMSE} = \sqrt{\frac{1}{N}\sum_{i=1}^{N} \left( E_{i}^\text{ref} - E_{i}^\text{model}\right)^2}.
\end{align}
This metric is sensitive to outliers, meaning that large errors
in a small number of samples can have a disproportionate impact
on the overall score.

On the other hand, MAD measures the average absolute difference
between predicted and actual values,
without taking into account the direction of the error
\begin{align}
    E^\text{MAD} = \frac{1}{N}\sum_{i=1}^{N} \left| E_{i}^\text{ref} - E_{i}^\text{model}\right|.
\end{align}
This metric is less sensitive to outliers, 
making it usually a good choice for cases with extreme values.
However when employing an MLP in simulations, bad predictions
for a small set of outliers can severely deteriorate the
quality of the sampling.
It is therefore usually more revealing to report
RMSE values rather than MAD.

Overall, both RMSE and MAD are commonly used in machine learning to validate
the performance of a model and to compare different models.
Using these metrics is a good starting point for validating a
machine learning model and ensure it is able to
accurately predict the reference data it was trained against.
The best practice is also to compute them for an independent validation
set rather than only for the training or test data.
This makes sure that no underlying bias in the selection of the training
data is skewing these performance metrics.
However, it is important to keep in mind that these measures
should be used in conjunction with other validation techniques
to ensure that machine learning models are performing as expected
and are able to accurately predict chemical properties.~\cite{Kovacs2021/10.1021/acs.jctc.1c00647}
Furthermore, users need to develop a feeling for the actual values of the error measures
for a given system and how they translate into actual performance in
simulations is not always clear.
\rev{
As a rule of thumb, energy errors below 1\,meV ($\approx$ 0.025\,kcal/mol
$\approx$ 0.1\,kJ/mol) per atom and force errors of 100\,meV/\AA{}
($\approx$ 2.5\,kcal/mol\AA{} $\approx$ 10\,kJ/mol\AA{}) or lower are usually
desirable.
Furthermore, the use of relative errors with respect to the
learned observable enables better comparison over the full
range of values and makes it easier to compare different systems.
}

\subsection{Validation of Secondary Properties}
\label{sec:validation_sec_prop}
While numerical errors can usually give a good initial assessment
of the quality of a developed MLP, it is important to validate
the prediction of the model more rigorously for the performance
in atomistic simulations.
After all, we want to use our model to predict physical quantities, and therefore other secondary properties derived from the
representation of the PES need to be tested with respect to
the reference method.
These are usually system and application-specific,
requiring some degree of domain knowledge.
It is therefore common to design a suite of validation
tests for the particular question at hand in order
to build trust in the predictions with the model.
Nevertheless, these tests can broadly be categorised
into structural and dynamical properties.

One example of a structural property that is commonly analysed
is radial distribution functions (RDFs).~\cite{Frenkel2001}
RDFs are defined for pairs of atom types and describe
how density varies as a function of distance from a reference particle.
\rev{
It is computed by counting the number of particles $dn_r$ within a shell of thickness $dr$
\begin{align}
g(r) = \frac{dn_r}{4\pi r^2 dr \cdot \rho^N}
\end{align}
divided by the spherical shell volume times the number density $\rho^N$.
}
$g(r)$ is related to many other static properties such as the
structure factor, or the potential of mean force
and thus gives a very good overview of the structural
properties of a system.
Given its pair-wise nature, it should only be considered
as the minimum condition in the validation of an MLP,
and it is usually important to check higher-order
structural properties such as angular or dihedral distributions.~\cite{Pinheiro2021/10.1039/d1sc03564a}

Dynamical properties include diffusion constants,
the phonon spectrum of solids, or more generally
the vibrational density of states (VDOS), as well as
IR or Raman response of a system of interest.
Given the additional complexity of predicting the latter
two observables due to the need for dipole moments and polarisabilities,
it is usually sufficient to evaluate the predictive power
of an MLP with respect to the simpler VDOS.
It can be obtained from the Fourier transform of
the velocity-velocity autocorrelation function and
can be readily dissected into atom-wise components
to give a more resolved overview.
\rev{
The frequency dependent VDOS $G_\alpha(\omega)$ for species $\alpha$ is then given by 
\begin{align}
 G_\alpha(\omega) =  \mathcal{F}(\langle v_\alpha(0)\cdot v_\alpha(t)\rangle),
\end{align}
where $\mathcal{F}$ denotes the Fourier transform of the autocorrelation function for velocities $v$ from time 0 to $t$ as ensemble average $\langle\cdots\rangle$.
}
Spanning the whole range of possible dynamical
processes in a system of interest from slow translational
and librational motion, up to bending and stretching modes,
the VDOS is a great summary of the performance of a model
for dynamical properties.
\rev{
Example code to obtain both RDFs and VDOS is available
in our Colab tutorial.
}

In many cases, the actual properties of interest for a system
under investigation are too expensive to be validated
explicitly with the reference method.
This is one of the main purposes of using machine learning
for atomistic simulations in order to push the boundaries
of what is doable with existing methods.
In such cases, sufficient trust in the model has to be
obtained based on cheaper and simpler properties.
Having access to an estimate of uncertainty during the
simulations, such as given by the above-mentioned committee
methods, or Gaussian process regression, can further help
to validate predictions for system sizes too large 
or simulation times too long to be treated with the reference method.

Finally, the prediction of a model can also be 
compared to experimental observables.
This is particularly suitable for cases where
the observable cannot be easily obtained with
the reference method, thus providing not only
an assessment of the model's performance, but
also of the underlying electronic structure method
and sampling technique.
At the same time, it does deviate slightly from
the pure assessment of the MLP, as factors such as
suitability of the reference method and approximations
in obtaining the observable also contribute to this
comparison.
Examples of this can be finite size effects in atomistic
simulations, limited statistics, but also underlying
approximations such as linear response theory.
Nevertheless, a comparison to the experiment can be understood
as the Holy Grail of validating simulations from first principles
and should always be part of a well-designed validation procedure.

\begin{figure*}[tb]
    \centering{}
    \includegraphics[width=1.0\textwidth]{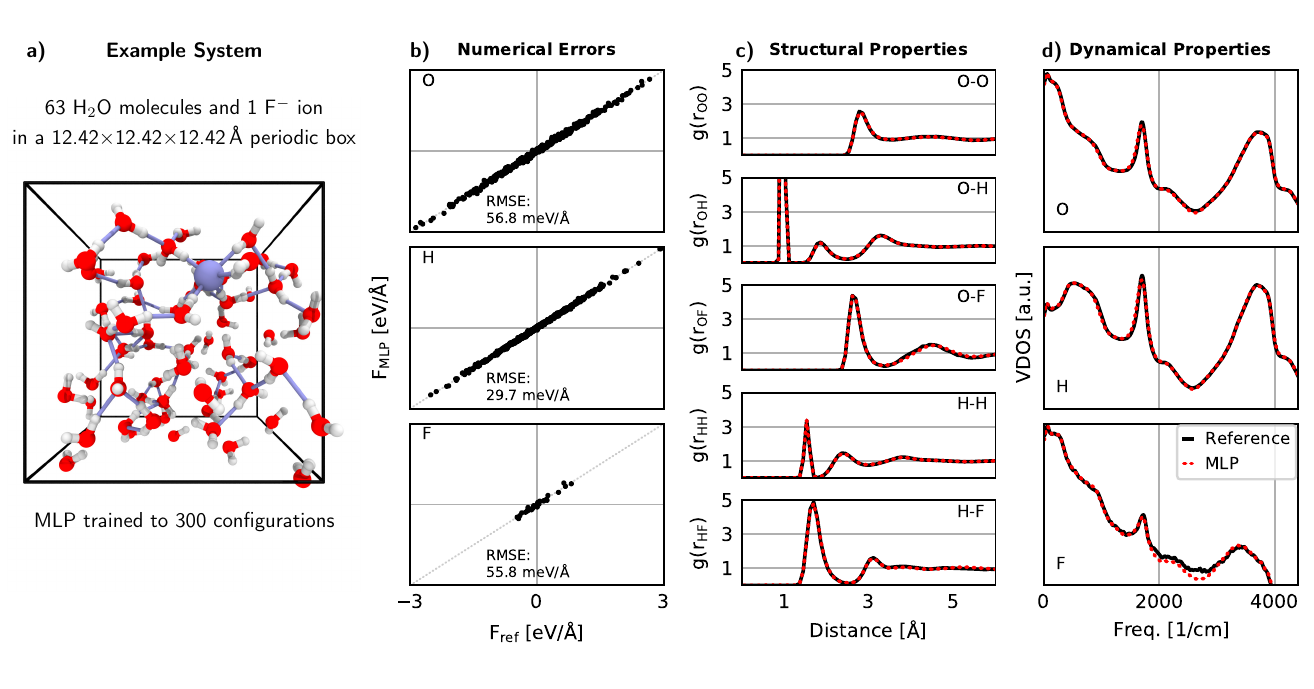}
    \caption{
    \textbf{Example of the validation of an MLP for the description of
    a fluoride ion in water.}
    a) Overview of the system of interest and trained MLP.
    b) Analysis of the numerical force errors of the MLP.
    The correlation of the reference force and prediction is shown
    for each element in the system together with the force RMSE.
    c) Performance of the MLP for structural properties encoded by the
    Radial Distribution Functions (RDFs) for all pairs of elements in the system.
    d) Performance of the MLP for dynamical properties as provided by
    the Vibrational Density of States (VDOS) for each element in the system.
    \rev{VDOS are shown in logarithmic scale to facilitate easier comparison over the full frequency range.}
    Figure adapted from Ref.~\citenum{Schran2021/10.1073/PNAS.2110077118}.
    }   
    \label{fig:validation}
\end{figure*}

An example for the validation of an MLP for primary and secondary properties
is shown in Figure~\ref{fig:validation}.
It is for a developed neural network based model of a solvated fluoride ion
by water as shown in panel a.
Numerical errors of the force prediction with respect to the reference
DFT method are given in panel b, while structural properties (RDFs)
and dynamical properties (VDOS) are validated in panels~c~and~d.
This example shows a well-developed model where reference and
prediction agree to a satisfactory degree.

\section{Showcase Examples}
\label{sec:examples}

In the last part of this tutorial, we will look at some showcase examples
that highlight what can be done with the machine learning techniques
discussed above.
We will give an overview of different applications that
rely either on kernel-based or neural network based MLPs
in combination with density functional theory and modern
sampling techniques.
\rev{
While impressive new developments with foreseeable high impact
are ongoing as described in detail in section~\ref{sec:devel},
we concentrate here on examples that highlight the transformative
power of MLPs to provide new scientific insight.
This has mostly been delivered by the two first established
techniques of HD-NNPs and GAP, which is why we dedicate a
larger proportion of the examples to these methods to best
showcase the state of the art of applications.
}

\begin{figure*}[th]
    \centering
    \includegraphics[width=\textwidth]{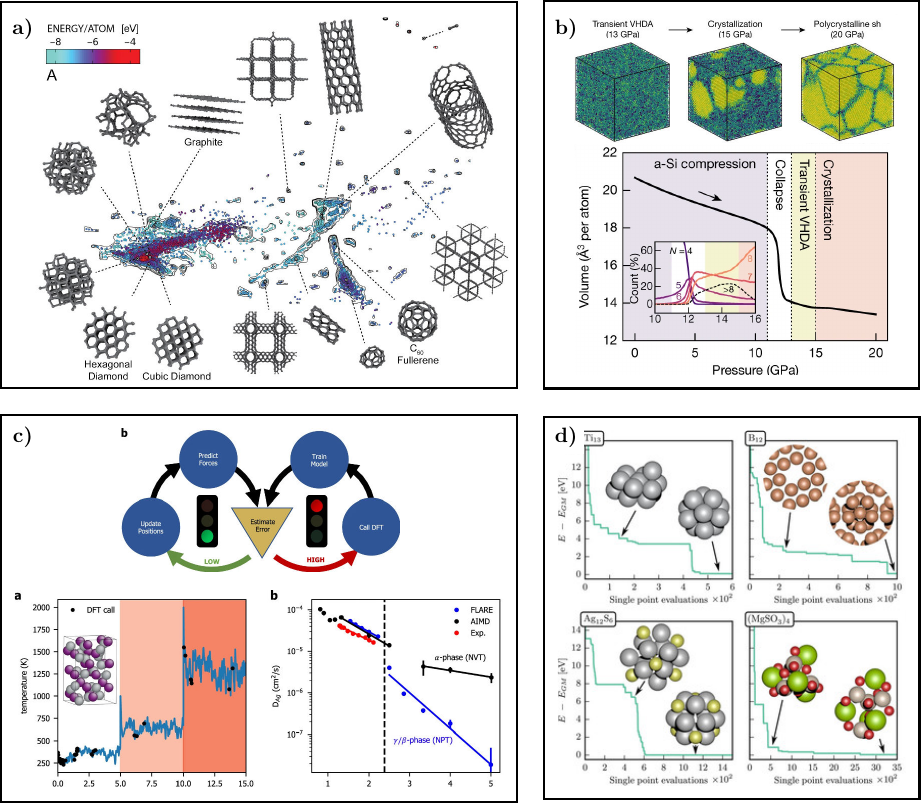}
    \caption{
        \textbf{Application of machine learning potentials using two different strategies.}
        The top row highlights use cases as general-purpose potentials for elementary systems
        such as carbon (a)~\cite{Rowe2020/10.1063/5.0005084} and silicon (b)~\cite{Deringer2021/10.1038/s41586-020-03072-z}.
        The bottom row features examples of surrogate models
        for on-the-fly learning (c)~\cite{Vandermause2020/10.1038/s41524-020-0283-z} and
        global structure optimisation~\cite{Ronne2022/10.1063/5.0121748}.
    \label{fig:strategy}
    }
\end{figure*}

In the last few years, GAP models have been very successful
in providing general-purpose potentials for elementary
systems.
The first set of examples highlights two cases for this particular application and
are given in Figure~\ref{fig:strategy}~a)~and~b).
The first one is a general-purpose model for carbon~\cite{Rowe2020/10.1063/5.0005084},
able to describe the rich allotropy of carbon
including diamond, graphene, graphite, nanotubes, and
fullerenes.
Furthermore, it is also applicable to amorphous 
and liquid phases, relevant for various technological
applications.
The model describes the relative stability of these
different polymorphs at a level of accuracy not achieved before
with force fields.
This has enabled studies of graphene rippling behaviour
and its dependence on defects in the material~\cite{Thiemann2021/10.1021/acs.nanolett.1c02585}.
The second is on work using a general purpose
GAP model of silicon to reveal new phase transition behaviour
when compressing silicon at high pressures.~\cite{Deringer2021/10.1038/s41586-020-03072-z}
Although not included in the training process of the model,
it was able to faithfully reproduce a previously unknown transient
phase observed before crystallisation.

\begin{figure*}[th]
    \centering
    \includegraphics[width=\textwidth]{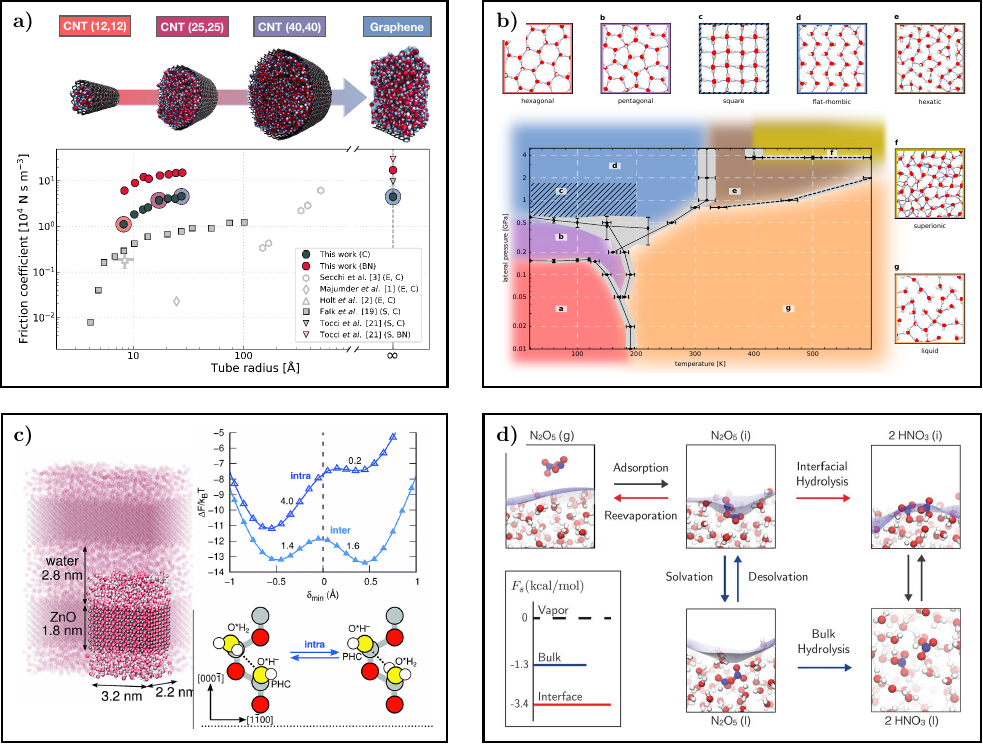}
    \caption{
        \textbf{Machine learning potentials applied to provide insight into complex aqueous systems.}
        a) Water flow in nanotubes~\cite{Thiemann2022/10.1021/acsnano.2c02784}.
        b) Phase behaviour of nanoconfined monolayer water~\cite{Kapil2022/10.1038/S41586-022-05036-X}.
        c) Water dissociation at a ZnO interface~\cite{Hellstrom2019/10.1039/c8sc03033b}.
        d) N$_2$O$_5$ decomposition at the water-air interface~\cite{Galib2021/10.1126/SCIENCE.ABD7716}.
    \label{fig:insight}
    }
\end{figure*}

The second set of examples at the bottom of Fig.\ref{fig:strategy} 
highlights another flavour of machine learning for atomistic simulations,
which is more tightly coupled with \textit{ab initio} sampling codes.
The general idea is that enabled by a robust uncertainty estimate in the MLP, it is possible
to train the model `on the fly' during sampling using a reference method
and switching to the model once it is accurate enough.
This drastically reduces the number of reference calculations and speeds up the sampling.~\cite{Csanyi2004/10.1103/PhysRevLett.93.175503, Li2015/10.1103/PhysRevLett.114.096405, Jinnouchi2019/10.1103/PhysRevB.100.014105, Pickard2022/10.1103/PHYSREVB.106.014102}
The bottom row of Fig.~\ref{fig:strategy} shows two applications of this approach, first to describe the phase behaviour
of silver iodide~\cite{Vandermause2020/10.1038/s41524-020-0283-z} (Fig.~\ref{fig:strategy} (c)), and second for finding global minima of various compounds~\cite{Ronne2022/10.1063/5.0121748},
while substantially reducing the number of required reference
calculations (Fig.~\ref{fig:strategy} (d)).
Another area of application of MLPs is in the modelling
of complex aqueous systems, as summarised in the third set of examples in Fig.~\ref{fig:insight}.
The first one is on water flow in different nanotubes~\cite{Thiemann2022/10.1021/acsnano.2c02784},
where experiments have reported interesting radius and material
dependence of the friction of water passing through.
This study was able to fully resolve the radius dependence
going to system sizes on the order of 10,000 atoms for multiple
nanoseconds simulation time.
Previous AIMD studies could only reach a few hundred picoseconds
for system sizes on the order of 500 atoms.
The second example shows the complex phase behaviour of
a single layer of water under nanoconfinement~\cite{Kapil2022/10.1038/S41586-022-05036-X}.
This setup can be realised experimentally by sandwiching
water in between graphene sheets.
This work revealed a rich phase diagram as a function of
temperature and pressure with two previously unreported
phases for this system:
A so-called hexatic phase, which is an intermediate between
solid and liquid, and a superionic phase with very high
propensity for proton transfer and thus high conductivity.
In this case, the use of MLPs has enabled the study of this system
at a level of complexity previously inaccessible by force fields and AIMD studies.

Finally, a big strength of MLPs over the traditional force field
approaches is the ability to describe bond breaking and formation.~\cite{Manzhos2021/10.1021/ACS.CHEMREV.0C00665, Meuwly2021/10.1021/ACS.CHEMREV.1C00033, Young2021/10.1039/D1SC01825F}
Some of the examples above have already shown this,
but the fourth set of examples in Fig.~\ref{fig:insight} highlights two studies that build more
on this capability.
The first is on the dissociation of water at the ZnO interface~\cite{Hellstrom2019/10.1039/c8sc03033b},
relevant for catalytic processes.
The second one shows the application of a neural network based model
to understand the decomposition of N$_2$O$_5$ at the water-air interface~\cite{Galib2021/10.1126/SCIENCE.ABD7716},
relevant for climate science.
Both of these studies provide insight into complex reactive processes
at interfaces, which would be very difficult or impossible to describe
using traditional approaches,
thus clearly highlighting how machine learning pushes forward
our ability to model complex processes with atomistic simulations.

Overall, these examples highlight the versatility of MLPs to provide insight into diverse areas of the natural sciences.
The general-purpose models for carbon and silicon fall primarily into the category of material science, while the silver iodide application is an example of condensed matter physics.
The applications to nanoconfined water showcase the ability of MLPs to study questions related to nanoscience.
At the same time, the last two on proton transfer reactions at interfaces belong to the field of acid-base chemistry, surface science, and catalysis.
The selected examples are only a tiny fraction of the vibrant field, and many other applications to various other areas of the natural sciences are published daily.
This use of a united set of tools over vastly different scientific areas is expected to continue.
We foresee a great future for MLPs in providing atomistic insight across the fields.

\section{Summary and Outlook}
\label{sec:outlook}
The integration of machine learning into the
representation of potential energy surfaces
in atomistic simulations can greatly
improve the accuracy and efficiency of these simulations.
Traditional methods for representing PESs rely on fitting analytical
functions to a limited set of data obtained from simulations or experiments.
However, these methods can be limited in their ability to accurately
represent the PESs for complex systems or materials.
ML algorithms, on the other hand, can learn the underlying
relationships between the atomic structure and the PES from a data-driven perspective, providing a more flexible and accurate representation of the PES.
Furthermore, MLPs can also be used for larger systems than used 
in the training process, without the need for additional simulations.
In some cases, even the application to previously unseen situations
can be achieved, thus relying on the transferability of the model~\cite{%
Rowe2020/10.1063/5.0005084,
Monserrat2020/10.1038/s41467-020-19606-y,
Schran2021/10.1063/5.0035438,
Zeni2022/10.1103/PhysRevB.105.165141}.
However, this should only be done with great care, as extrapolation
does, in general, not work with machine learning models.
\rev{
The first introduced methods HD-NNPs, GAP, and others have delivered
great scientific insight and have paved the way for new developments
which further transform the field of molecular and materials modelling.
}

\rev{
Let us summarise the relevant concepts behind MLPs once more.
First, a meaningful set of structures has to be curated, for example, with a force field or ab initio sampling in the initial stage.
The coordinates of that initial training set are then transformed by a set of descriptors that incorporate the relevant invariances.
Next, these are used as input for the regression model of choice, outputting atomic energy contributions for each atom in the system, summing up to the total potential energy.
The model's parameters are optimised by comparing the predicted energy (and usually the forces) to the reference values from the ab initio method of choice.
Validation of important properties will show if this initial model is sufficient for the envisaged application.
If not, the training set is expanded either by structure-based or property-based selection techniques (or both) and the process is repeated until a satisfactory quality is achieved.
Once an initial model is available, new structures can readily be generated with that model, usually speeding up the exploration process.
Meaningful error estimates can benefit this step as they enable the identification of outliers and validation of simulation results.
Finally, the developed model passes all tests relevant to the application and can be applied to provide insight into challenging scientific questions.
}

The data-driven and automated approaches to develop new MLPs and select training data
have significantly reduced the required number of reference calculations.
This enables sophisticated electronic structure calculations 
to be used as a reference for the model, which would otherwise
be too expensive for on-the-fly sampling.
Examples include the use of converged coupled-cluster calculations
for gas phase systems such as reactive protonated water clusters~\cite{Schran2018/10.1063/1.4996819,
Schran2020/10.1021/acs.jctc.9b00805},
organic molecules~\cite{Chmiela2018/10.1038/s41467-018-06169-2}, and
even models over chemical compound space.~\cite{Smith2019/10.1038/s41467-019-10827-4}
Furthermore, it has been shown recently that these techniques can
also be leveraged for condensed phase systems, as demonstrated for
liquid water at coupled cluster accuracy~\cite{Daru2022/10.1103/PhysRevLett.129.226001,
Chen2023/10.1021/acs.jctc.2c01203},
or high-pressure phases of hydrogen using variational Monte Carlo~\cite{Cheng2020/10.1038/s41586-020-2677-y}.
Some of these examples make use of techniques to further reduce the number of reference points,
such as delta learning~\cite{Ramakrishnan2015/10.1021/acs.jctc.5b00099},
or transfer learning~\cite{Smith2019/10.1038/s41467-019-10827-4,
Zaverkin2023/10.1039/d2cp05793j}.
The former represents only the difference between a high-level and a low-level
method via machine learning, thus making the resulting delta-PES
smoother and easier to learn.
The latter uses a pre-trained model optimized to a large set of low-level reference points
and retrains to a much smaller set of high-level points, thus transferring parts of the
learned physics of the PES from the cheaper to the more demanding method.
This push for high-quality reference methods is very promising
and expected to flourish in the following years,
opening up the possibility of describing yet more challenging systems at
previously unattainable accuracy.
Another area where ML has the potential to improve and accelerate atomistic 
simulations is in the representation of other properties,
such as dipole moments \cite{Gastegger2017/10.1039/C7SC02267K, Litman2019/10.1039/C9FD00056A}, or polarisabilities \cite{Wilkins2019/10.1073/PNAS.1816132116}.
These properties are often difficult or expensive to calculate using
traditional methods but can be equally well learned directly from reference data.
They play a crucial role in determining the optical and
electronic properties of materials, and their accurate prediction is
essential for the rational design and optimisation of materials for technological applications.
This also includes excited potential energy surfaces, enabling the treatment of
electron excitation processes and excited state dynamics.~\cite{Westermayr2020/10.1088/2632-2153/AB88D0, Westermayr2021/10.1021/ACS.CHEMREV.0C00749}
Representing wave functions~\cite{Schutt2019/10.1038/s41467-019-12875-2} or electronic densities~\cite{Grisafi2019/10.1021/ACSCENTSCI.8B00551} with ML is another
active area of research, which has the potential to greatly
improve the accuracy of quantum mechanical simulations.
In these cases, rather than learning derived properties, the wave function
is learned directly enabling all derived properties to be easily calculated. 

One of the remaining challenges in representing PESs with ML is the long-range
interactions between atoms~\cite{Behler2021/10.1140/EPJB/S10051-021-00156-1}.
These interactions can have a significant impact on the properties of
materials but require some degree of physics to be included in the ML model.
\rev{
Systems that are particularly impacted by short-sighted models
include polar crystal surfaces~\cite{Cox2022/10.1063/5.0097531,Quaranta2019/10.1021/ACS.JPCC.8B10781},
disordered interfaces such as the water-air interface~\cite{Cox2020/10.1073/pnas.2005847117,Niblett2021/10.1063/5.0067565},
and systems with long-range charge transfer~\cite{Ko2021/10.1038/s41467-020-20427-2}.
}
Recent work has shown that ML can be used to accurately represent
these long-range interactions by using different techniques and
this push towards more physics-inspired models is expected to continue.~\cite{Gao2022/10.1038/s41467-022-29243-2}
These methods can accurately predict the PESs for
a wide range of materials where long-range interactions are important,
including disordered or polar interfaces.~\cite{Quaranta2019/10.1021/ACS.JPCC.8B10781, Artrith2019/10.1088/2515-7655/ab2060, Niblett2021/10.1063/5.0067565}

Finally, the recent serge in the development of generalisable models, such as
MACE-MP-0~\cite{batatia2023foundation}, MACE-OFF~\cite{kovacs2023maceoff23}, GNoME~\cite{Merchant2023/10.1038/s41586-023-06735-9}, MatterSim~\cite{yang2024mattersimdeeplearningatomistic}, CHGNet\cite{deng2023chgnet}, and others,
highlights the potential for MLPs to deliver universal force fields applicable
across chemical compound space.
Their development and push towards more robustness and accuracy has only begone,
but shows great promises even beyond the regimes set by their training data.
Additional advantage of these models is in providing a starting point as structure
generator and for fine tuning according to specific needs of an application of interest
in terms of chemical composition and electronic structure reference.
Out of the many recent developments, this has lead to exciting progress and will open up these tools to an even wider community, including non-expert users.

In conclusion, ML is revolutionising the way we represent
and predict the properties of materials and reactions by atomistic simulations.
The integration of ML into atomistic simulations has shown great promise
in the representation of interatomic potential energy surfaces,
prediction of other properties such as dipole moments, polarisabilities,
and excited states.
Representing wave functions or electronic densities with ML is also
a promising area of research.
There are still many open questions and challenges to be addressed,
such as the long-range interactions, generalisation and interpretability.
Nevertheless, machine learning for atomistic simulations has proven to
be a game changer in the field, providing a new understanding of complex systems.
It is here to stay and will continue to deliver exciting new approaches that
make it possible to tackle more and more complex and challenging scientific
problems.

\begin{acknowledgments}
We would like to thank Gabor Csanyi, J\"org Behler, Ondrej Marsalek, and Dominik Marx
for many discussions on this topic.
C.S. acknowledges partial financial support from the
\textit{Alexander von Humboldt-Stiftung} and the Deutsche Forschungsgemeinschaft (DFG, German Research Foundation) project number 500244608. 
N.O.N acknowledges financial support from the Gates Cambridge Trust.
V.K. acknowledges support from the Ernest Oppenheimer Early Career Fellowship and the Sydney Harvey Junior Research Fellowship, Churchill College, University of Cambridge.
AM acknowledges support from the European Union under the ``n-aqua" ERC project (101071937).
\end{acknowledgments}
\section*{Competing interests}
The authors declare no competing interests.

\section*{Supporting Information}
\rev{
To better facilitate the understanding of the described concepts,
we have developed a Colab online tutorial that walks users
through all relevant steps of developing an MLP.
It is focused on a simple, one component system, diamond, for which a training set is constructed using query by committee from a short reference simulation.
Next, the resulting model is used for a longer simulation and validated with respect to the reference.
This tutorial can be accessed via \href{https://colab.research.google.com/drive/19wQ3t8Lo97Ul1HYrJ5GQS2XtZxEBNd86?usp=sharing}{Colab}.
}
An overview of \rev{the most important ML concepts is provided in Tab.\ref{tab:def}, while} different open-source codes to develop machine learning potentials
\rev{are compiled} in Tab.~\ref{tab:code} including links to the software packages
as well as relevant citations~\cite{Behler2007/10.1103/PhysRevLett.98.146401,
Singraber2019/10.1021/acs.jctc.8b01092,
Unke2019/10.1021/ACS.JCTC.9B00181,
Schutt2017,
Zeng2023/2304.09409,
Batatia2022/10.48550/arxiv.2206.07697,Batatia2022Design,
Batzner2021/10.1038/s41467-022-29939-5,thomas_tensor_2018,geiger_e3nn_2022,
Fan2022/10.1063/5.0106617,
Gao2020/10.1021/ACS.JCIM.0C00451,
Artrith2016/10.1016/J.COMMATSCI.2015.11.047,
Lot2020/10.1016/J.CPC.2020.107402,
Bartok2010,
Bochkarev2022/10.1103/PhysRevMaterials.6.013804,Lysogorskiy2021/10.1038/s41524-021-00559-9,Drautz2019/10.1103/PHYSREVB.99.014104,
Vandermause2020/10.1038/s41524-020-0283-z,
chmiela2017/10.1126/SCIADV.1603015,
Shapeev2016/10.1137/15M1054183,
Christensen2020/10.1063/1.5126701/1064737,
Bisbo2020/10.1103/PhysRevLett.124.086102,
Hajibabaei2021/10.1103/PhysRevB.103.214102}.

%
%

%

%

\clearpage

\setcounter{section}{0}
\setcounter{equation}{0}
\setcounter{figure}{0}
\setcounter{table}{0}
\setcounter{page}{1}

\renewcommand{\thesection}{S\arabic{section}}
\renewcommand{\theequation}{S\arabic{equation}}
\renewcommand{\thefigure}{S\arabic{figure}}
\renewcommand{\thetable}{S\arabic{table}}
\renewcommand{\thepage}{S\arabic{page}}
\title{Supporting information for: \mytitle}
{\maketitle}

\onecolumngrid

\section*{Overview of Machine Learning Concepts}
\label{sec:def}
\begin{table*}[th!]
    \caption{%
    \textbf{Definition of important concepts.}
\label{tab:def}}
\begin{tabularx}{\textwidth}{p{0.25\linewidth}|X}
\hline
\textbf{Keyword} & \textbf{Definition} \\
\hline
Machine learning & Draw inferences from patterns in data using algorithms\\
Supervised learning & Input data (structures) is provided by user and a desired output (energy, force) is learned by ML model.\\
ML model & Universal function with many parameters to be optimised\\
Regression & Transform structure to energy\\
Training set & Data used for optimisation \\
Test set & Data used to check transferability of model \\
Label & Reference energy (and forces) from electronic structure calculation \\
Hyperparameters & Parameters not optimised during learning task, but chosen by user (number of iterations in optimisation, size of model, ...)\\
\hline
\end{tabularx}
\end{table*}

\FloatBarrier
\newpage
\section*{Overview of Open-Source Code}
\label{sec:open-source}
\begin{table*}[th]
\caption{%
    \textbf{Overview of different open-source codes to develop machine learning potentials.}
\label{tab:code}}
\begin{tabularx}{\textwidth}{p{0.25\linewidth}|X}
\hline
\multicolumn{1}{l}{\textbf{Code}} & \multicolumn{1}{l}{\textbf{Description}} \\ \hline
\multicolumn{2}{c}{\textbf{HD-NNP}} \\ \hline
\href{https://theochemgoettingen.gitlab.io/RuNNer}{RuNNer}~\cite{Behler2007/10.1103/PhysRevLett.98.146401}  & Original implementation of HD-NNPs. \\
\href{https://github.com/CompPhysVienna/n2p2}{n2p2}~\cite{Singraber2019/10.1021/acs.jctc.8b01092}  & Library based modular implementation of HD-NNPs with interface to Lammps. \\
\href{https://github.com/MMunibas/PhysNet}{PhysNet}~\cite{Unke2019/10.1021/ACS.JCTC.9B00181} & Tensorflow based message-passing NNP implementation built on physical principles for predicting energies, forces, dipole moments and partial charges. \\
\href{https://github.com/atomistic-machine-learning/schnetpack}{SchNetPack}~\cite{Schutt2017} & Message-passing NN model based on pairwise distances with GPU accelerated MD code and output modules for dipole moment, polarizability, stress etc.  \\
\href{https://github.com/deepmodeling/deepmd-kit}{DeepMD}~\cite{Zeng2023/2304.09409} &  Deep NN model based on three body correlation functions with possibility of learning tensors. \\
\href{https://github.com/ACEsuit/mace}{MACE}~\cite{Batatia2022/10.48550/arxiv.2206.07697, Batatia2022Design} &  Interatomic potentials with higher order equivariant message passing and $\mathcal{O}(1)$ scaling with number of chemical species.\\
\href{https://github.com/mir-group/nequip}{Nequip}~\cite{Batzner2021/10.1038/s41467-022-29939-5} & Based on tensor field networks~\cite{thomas_tensor_2018} implemented in e3nn~\cite{geiger_e3nn_2022}, a general framework for building E(3)-equivariant neural networks.\\
\href{https://gpumd.org/introduction.html}{GPUMD}~\cite{Fan2022/10.1063/5.0106617} & GPU accelerated molecular dynamics code that supports neuroevolution potentials. \\
\href{https://github.com/aiqm/torchani}{TorchANI}~\cite{Gao2020/10.1021/ACS.JCIM.0C00451} & PyTorch-based implementation of the ANI NNP.\\
\href{http://ann.atomistic.net/}{aenet}~\cite{Artrith2016/10.1016/J.COMMATSCI.2015.11.047} & Training code for NNPs with interface to Tinker. \\
\href{https://gitlab.com/PANNAdevs/panna}{PANNA}~\cite{Lot2020/10.1016/J.CPC.2020.107402} & TensorFlow based package to train and validate NNPs with lammps and ase interface.\\ \hline
\multicolumn{2}{c}{\textbf{Kernel-Based MLPs}} \\ \hline
\href{https://github.com/libAtoms/QUIP}{QUIP}~\cite{Bartok2010} & Original implementation of the Gaussian Approximation Potential \\
\href{https://github.com/ICAMS/python-ace}{pacemaker}~\cite{Bochkarev2022/10.1103/PhysRevMaterials.6.013804,Lysogorskiy2021/10.1038/s41524-021-00559-9,Drautz2019/10.1103/PHYSREVB.99.014104} &  Tool for fitting of interatomic potentials in a general nonlinear Atomic Cluster Expansion form. \\
\href{https://github.com/ACEsuit}{ACEsuit}~\cite{Lysogorskiy2021/10.1038/s41524-021-00559-9} & Various software packages surrounding the atomic cluster expansion written in Julia. \\
\href{https://github.com/mir-group/flare}{flare}~\cite{Vandermause2020/10.1038/s41524-020-0283-z} & ACE descriptors coupled to sparse GP used for on-the-fly learning and interfaced to Lammps.  \\
\href{https://github.com/stefanch/sGDML}{sGDML}~\cite{chmiela2017/10.1126/SCIADV.1603015} & Symmetric Gradient Domain Machine Learning implementation.\\ 
\href{https://gitlab.com/ashapeev/mlip-2}{MTP}~\cite{Shapeev2016/10.1137/15M1054183} &  Original implementation of the Moment Tensor Potential. \\
\href{https://github.com/qmlcode/qml}{FCHL}~\cite{Christensen2020/10.1063/1.5126701/1064737} &  MLP based on distribution functions of structural and alchemical parameters of atoms. \\
\href{http://grendel-www.cscaa.dk/mkb/}{GOFEE}~\cite{Bisbo2020/10.1103/PhysRevLett.124.086102} & Efficient global structure optimization with a machine-learned surrogate model.\\
\href{https://github.com/amirhajibabaei/AutoForce}{AutoForce}~\cite{Hajibabaei2021/10.1103/PhysRevB.103.214102} & Python package for sparse GPR of ab-initio PES.\\
\hline
\end{tabularx}
\end{table*}

\end{document}